\documentclass[prd,twocolumn,twoside,preprintnumbers,superscriptaddress,floatfix,reprint]{revtex4}
\usepackage[utf8]{inputenc}
\usepackage{amsmath,slashed}
\usepackage{amsfonts}
\usepackage{graphicx,graphics,color}
\usepackage{dcolumn}
\usepackage{xcolor} 
\definecolor{solarized1}{HTML}{859900}
\definecolor{solarized2}{HTML}{2AA198}
\usepackage[hyperfootnotes=false]{hyperref}
\hypersetup{
  colorlinks=true,
  urlcolor     = solarized2, 
  linkcolor    = solarized2, 
  citecolor   = solarized1 
}
\usepackage{xspace}
\usepackage{siunitx}
\usepackage{pgfplots}
\pgfplotsset{compat=1.13} 
\usepackage[tight]{subfigure} 

\DeclareUnicodeCharacter{2212}{-} 

\let\originalleft\left
\let\originalright\right
\renewcommand{\left}{\mathopen{}\mathclose\bgroup\originalleft}
\renewcommand{\right}{\aftergroup\egroup\originalright}

\newcommand{\Eqref}[1]{Eq.~\eqref{#1}}
\newcommand{\Figref}[1]{Fig.~\ref{#1}}

\newcommand{\mc}[1]{\mathcal{#1}}

\newcommand{\mb}[1]{\mathbb{#1}}

\newcommand{\Fpi}{F_\pi}
\newcommand{\mpi}{M_{\pi}}
\newcommand{\mrho}{M_{\rho}}

\newcommand{\beq}{\begin{equation}}
\newcommand{\eeq}{\end{equation}}
\newcommand{\Li}{\text{Li}}
\renewcommand{\Re}{\text{Re}\,}
\renewcommand{\Im}{\text{Im}\,}

\renewcommand{\vec}[1]{\boldsymbol{#1}}

\allowdisplaybreaks[1]

\newcommand{\thline}{\hline \\[-3mm]}

\begin{document}

\preprint{INT-20-034}
\title{Two-loop analysis of the pion-mass dependence of the $\boldsymbol{\rho}$ meson}

\author{Malwin Niehus}
\email{niehus@hiskp.uni-bonn.de}
\affiliation{Helmholtz-Institut f\"ur Strahlen- und Kernphysik (Theorie) and\\
Bethe Center for Theoretical Physics, Universit\"at Bonn, 53115 Bonn, Germany}

\author{Martin Hoferichter}
\email{hoferichter@itp.unibe.ch}
\affiliation{Albert Einstein Center for Fundamental Physics, Institute for Theoretical Physics, University of Bern, Sidlerstrasse 5, 3012 Bern, Switzerland}
\affiliation{Institute for Nuclear Theory, University of Washington, Seattle, WA 98195-1550, USA}

\author{Bastian Kubis}
\email{kubis@hiskp.uni-bonn.de}
\affiliation{Helmholtz-Institut f\"ur Strahlen- und Kernphysik (Theorie) and\\
Bethe Center for Theoretical Physics, Universit\"at Bonn, 53115 Bonn, Germany}

\author{Jacobo Ruiz de Elvira}
\email{elvira@itp.unibe.ch}
\affiliation{Albert Einstein Center for Fundamental Physics, Institute for Theoretical Physics, University of Bern, Sidlerstrasse 5, 3012 Bern, Switzerland}

\begin{abstract}
Analyzing the pion-mass dependence of $\pi\pi$ scattering phase shifts beyond the low-energy region requires the unitarization of the amplitudes from chiral perturbation theory. In the two-flavor theory, unitarization via the inverse-amplitude method (IAM) can be justified from dispersion relations, which is therefore expected to provide reliable predictions for the pion-mass dependence of results from lattice QCD calculations. In this work, we provide compact analytic expression for the two-loop partial-wave amplitudes for $J=0,1,2$ required for the IAM at subleading order. To analyze the pion-mass dependence of recent lattice QCD results for the $P$-wave, we develop a fit strategy that for the first time allows us to perform stable two-loop IAM fits and assess the chiral convergence of the IAM approach. While the comparison of subsequent orders suggests a breakdown scale not much below the $\rho$ mass, 
a detailed understanding of the systematic uncertainties of lattice QCD data is critical to obtain acceptable fits, especially at larger pion masses.   
\end{abstract}

\maketitle

\section{Introduction}
While recent years have shown significant progress in understanding the QCD resonance spectrum from first principles in lattice QCD~\cite{Briceo2018}, most calculations are still performed at unphysically large pion masses, requiring an extrapolation to the physical point to make connection with experiment. Such extrapolations can be controlled using effective field theories, i.e., chiral perturbation theory (ChPT)~\cite{Weinberg:1978kz,Gasser1984,Gasser:1984gg} for observables that allow for a perturbative expansion. By definition, this precludes a direct application to resonances such as the $\rho$ meson in the $P$-wave of $\pi\pi$ scattering. 
In fact, spectroscopy results from lattice QCD are arguably most
advanced for the $\rho$ meson~\cite{Aoki:2007rd,Gockeler:2008kc,Feng:2010es,Lang:2011mn,Pelissier:2012pi,Metivet:2014bga,Feng:2014gba,Wilson2015,Bali:2015gji,Guo:2016zos,Fu:2016itp,Alexandrou:2017mpi,Andersen2019,Werner:2019hxc,Erben:2019nmx,Fischer:2020fvl}, with even
calculations at the physical point now available~\cite{Fischer:2020fvl}, which
makes this channel the ideal example to study the details of the pion-mass dependence. In addition, the $\pi\pi$ $P$-wave features prominently
in a host of phenomenological applications, among them hadronic vacuum polarization~\cite{Aoyama:2020ynm,Colangelo:2018mtw,Ananthanarayan:2018nyx,Davier:2019can,keshavarzi:2019abf,Colangelo:2020lcg}, nucleon form factors~\cite{Belushkin:2006qa,Lorenz:2014yda,Hoferichter:2016duk,Hoferichter:2018zwu}, and the radiative process $\gamma\pi\to\pi\pi$~\cite{Hoferichter:2012pm,
Hoferichter:2017ftn}. For the latter, a thorough understanding of the $\pi\pi$ $P$-wave is prerequisite for an analysis of the pion-mass dependence of recent lattice results~\cite{Briceno:2015dca,Briceno:2016kkp,Alexandrou:2018jbt}, see Ref.~\cite{Niehus:2019nkl}, and similarly for decays into three-pion final states~\cite{Dax:2018rvs}.

On the technical level, the failure to produce resonant states is related to the fact that unitarity is only restored perturbatively in ChPT, so that any description of resonances requires a unitarization procedure. A widely used approach known as the IAM achieves this unitarization by studying the unitarity relation for the inverse amplitude~\cite{Truong:1988zp,Dobado:1989qm,Truong1991,Dobado1993,Dobado1997,Guerrero:1998ei,GomezNicola:2001as,Nieves:2001de,Dobado:2001rv}. In particular, in the case of $SU(2)$ ChPT the IAM procedure can be derived starting from a dispersion relation in which the discontinuity of the left-hand cut is approximated by its chiral expansion~\cite{Dobado1993,Dobado1997}. While Adler zeros induce a modification for the $S$-waves~\cite{GomezNicola:2007qj}, the naive derivation of the IAM survives for the $P$-wave amplitude: writing the partial wave for $\pi\pi$ scattering $t(s)$ as
\begin{equation} \label{eq:chpt_expansion}
    t\left(s\right) = t_2\left(s\right) + t_4\left(s\right) + t_6\left(s\right),
\end{equation}
with the subscripts indicating the chiral order, the unitarized amplitude at next-to-leading order (NLO) becomes~\cite{Dobado:1989qm,Truong1991,Dobado1993}
\begin{equation}\label{eq:nlo_iam}
    t_{\text{NLO}}\left(s\right) = \frac{\big[t_2\left(s\right)\big]^2}{t_2\left(s\right) - t_4\left(s\right)},
\end{equation}
while at next-to-next-to-leading order (NNLO)~\cite{Dobado1997,Nieves:2001de}
\begin{equation}\label{eq:nnlo_iam}
    t_{\text{NNLO}}\left(s\right) = \frac{\big[t_2\left(s\right)\big]^2}{t_2\left(s\right) - t_4\left(s\right) + \big[t_4\left(s\right)\big]^2/t_2\left(s\right) - t_6\left(s\right)}.
\end{equation}
To assess the chiral expansion of the unitarized amplitude beyond the first term, one thus needs the partial-wave amplitudes at two-loop order~\cite{Bijnens1996}.  

The IAM has been applied to study resonance properties at unphysical pion masses at one- and two-loop order as early as in Refs.~\cite{Hanhart:2008mx,Nebreda:2010wv,Pelaez:2010fj,Nebreda:2011di}, with numerous subsequent works confronting the IAM predictions with lattice data~\cite{Bolton:2015psa,Hu:2016shf,Doring:2016bdr,Hu:2017wli,Molina:2020qpw}. However, apart from Refs.~\cite{Pelaez:2010fj,Nebreda:2011di} such studies have been restricted to one-loop order, so that it was not possible to scrutinize the convergence properties of the expansion in the  pion mass.         

The reason for this situation was twofold: first, while the one-loop amplitudes can be given in analytic form, similarly compact expressions were not available for the two-loop amplitudes, thus complicating their implementation considerably. Second, as shown in Refs.~\cite{Pelaez:2010fj,Nebreda:2011di}, the increased number of low-energy constants (LECs) renders the fits more volatile, so that lattice data need to reach a sufficient quality to allow for meaningful two-loop fits. In this Letter we address both points: we present compact analytic expressions for the two-loop amplitudes that are straightforward to implement and devise a strategy for stable two-loop fits to current lattice data. While expressions are provided for all partial waves up to $J=2$, we concentrate on the application to the $\pi\pi$ $P$-wave, including the resonance parameters of the $\rho$ meson and its pole residue.      
 
\section{Partial waves in ChPT}
We express the partial waves $t_J^I(s)$, where $I$ and $J$ stand for the isospin and angular momentum, respectively, in terms of the pion decay constant in the chiral limit $F$ as well as the pion mass $\mpi$ (including quark-mass corrections from the LEC $l_3^r$), to render the dependence on the physical pion mass explicit and exclude a spurious mass dependence arising from the transition $F\to F_\pi$~\cite{Supp}\nocite{2020SciPy-NMeth,Edwards:2011jj}. We will follow the conventions of Refs.~\cite{Gasser1984,Bijnens:1999hw} for the one-loop LECs $l_i^r$ and the two-loop LECs $r_i^r$.
First, the leading-order (LO) results are~\cite{Weinberg:1966kf}
\begin{align}
t_0^0(s)\big|_2&=\frac{2s-\mpi^2}{32\pi F^2},\qquad t_0^2(s)\big|_2=-\frac{s-2\mpi^2}{32\pi F^2},\notag\\ 
t_1^1(s)\big|_2&= \frac{s-4\mpi^2}{96\pi F^2},\qquad t_2^I(s)\big|_2=0.
\end{align}
At NLO, the partial-wave amplitudes can be written in the form  
\beq
\Re t_J^I(s)\big|_4= \sum_{i=0}^2 b_i^{IJ}(s) \left[L(s)\right]^i +  \sum_{i=1}^3 b_{l_i}^{IJ}(s) l_i^r,
\eeq
in terms of 
\beq
L(s)=\log\frac{1+\sigma(s)}{1-\sigma(s)},\qquad \sigma(s)=\sqrt{1-\frac{4\mpi^2}{s}},
\eeq 
and coefficient functions $b_i^{IJ}(s)$, $b_{l_i}^{IJ}(s)$, which apart from phase-space and angular-momentum factors are polynomials in $s$.
The explicit expressions are listed in App.~\ref{sec:ChPT}.
We find that the NNLO expressions can be brought into a very similar form   
\begin{align}
  \Re t_J^I(s)\big|_6&=\sum_{i=0}^4c_i^{IJ}(s) \left[L(s)\right]^i+ \sum_{i=1}^3c_{l_i}^{IJ}(s) l_i^r \notag\\
 &+d^{Il}(s)\bigg[\sum_{n=\pm}\Li_3\big(\sigma_n(s)\big)-L(s)\, \Li_2\big(\sigma_-(s)\big)\Big]\notag\\
 &+c_{l_3^2}^{IJ}(s)\big(l_3^r\big)^2+P^{IJ}(s), 
\end{align}
where $\sigma_\pm(s)=2\sigma(s)/(\sigma(s)\pm 1)$ and 
in addition to powers of $L(s)$ also polylogarithms $\Li_n$ appear. 
The contributions from the NNLO LECs are collected in $P^{IJ}(s)$ and the imaginary parts determined by perturbative unitarity  
\beq
\Im t_4(s)=\sigma(s) \big[t_2(s)\big]^2,\quad
\Im t_6(s)=2\sigma(s)\, t_2(s) \Re t_4(s). 
\eeq

\section{Fits to lattice data}
From here on, we focus on the $P$-wave of $\pi\pi$ scattering, with both isospin $I$ and angular momentum $J$ equal to one.
Its phase $\delta(s) = \arg(t_1^1(s))$ can be computed using lattice QCD via L\"{u}scher's quantization condition~\cite{Luscher:1990ux,Briceo2018}, which allows one to determine the phase shift given $\pi\pi$ energy levels and vice versa.
To illustrate the fitting strategy as well as the conclusions regarding the pion-mass dependence of $\delta$ and the $\rho$ parameters, we analyze such energy levels as computed on the lattice by two different groups:
first, the one from Ref.~\cite{Andersen2019}, based on gauge configurations generated by the CLS collaboration, accompanied by a determination of the pion decay constant~\cite{Bruno:2016plf}.
There are six data sets (ensembles) at five different pion masses in the range \SIrange{200}{284}{\MeV}.
Second, we consider the energy levels from the HadSpec collaboration~\cite{Dudek2013,Wilson2015}, using one of their ensembles with $\mpi\approx\SI{236}{\MeV}$ and two with $\mpi\approx\SI{391}{\MeV}$.
Both lattice calculations involve $N_\text{f} = 2+1$ flavor simulations, but in either case the changes compared to the physical kaon mass, which determine the corrections to the LECs in two-flavor ChPT~\cite{Gasser:2007sg,Gasser:2009hr}, are negligibly small compared to other sources of uncertainty.
In the following, we concentrate mainly on the fit to the CLS data; a detailed description of the fitting procedure as well as an overview over the lattice data is given in App.~\ref{sec:appendix_lattice_details}, while the fit to the HadSpec data is discussed in App.~\ref{sec:appendix_HadSpec_results}.
To reduce the impact of scale-setting uncertainties, i.e., the error that arises when determining the lattice spacing in physical units, we work in lattice units wherever possible.
The fit proceeds as follows:
at NLO, Eq.~\eqref{eq:nlo_iam} is used to compute the phase $\delta$, which is subsequently inserted into L\"{u}scher's quantization condition to determine the energy levels.
Their distance to the energies as computed on the lattice is then minimized.
Simultaneously, the pion decay constant is fit, using the ChPT expression given in App.~\ref{sec:appendix_pion_decay_chpt}, truncated at NLO.
In an NNLO fit, the same procedure is applied, with Eq.~\eqref{eq:nnlo_iam} instead of Eq.~\eqref{eq:nlo_iam} and the pion decay constant truncated at NNLO.
This means that at NLO only the LECs $l_2^r - 2l_1^r$ and $l_4^r$ appear, while the NNLO expressions depend on $l_{1\text{--}4}^r$ as well as $r_{a,b,c}$ and $r_F^r$.
The minimization of the $\chi^2$ with respect to the fit parameters---most importantly the LECs---requires a sufficiently powerful algorithm.
To find the global minimum, we first employ the differential evolution algorithm~\cite{Storn1997}, whose results are subsequently refined via a modification of Powell's method~\cite{Press:2007num}.
The former algorithm allows one to tackle the multi-dimensional, non-linear optimization problem at hand in both a robust and efficient manner, if its parameters are adjusted carefully.
Together with the improved lattice data the choice and tuning of this algorithm are crucial to obtain sound fits that are stable even when ensembles at only a few different pion masses are available, e.g., the two masses used by HadSpec.
There are three sources of error that need to be considered for a reliable uncertainty estimate.
First, the statistical error of the lattice data.
Second, the error of the lattice spacing, which enters the ChPT expressions indirectly via the renormalization scale $\mu$, see App.~\ref{sec:appendix_error}.
Third, the error that arises as a result of the truncation of the chiral expansion~\eqref{eq:chpt_expansion}, which we are able to study in detail by a comparison of the IAM at one- and two-loop order. 
The chiral expansion proceeds in $s / \mrho^2$ as well as $\alpha = \mpi^2 / \mrho^2$, with the breakdown scale expected to be set by the $\rho$ mass since it is the lowest-lying resonance in the partial wave of interest. The energy dependence is resummed by the unitarization via the IAM, leaving the expansion in the pion mass as the most critical variable. 
Following Ref.~\cite{Epelbaum:2014efa}, we estimate the truncation error of an observable $X$ as
\begin{align}
    \Delta X_\text{NLO} &= \alpha X_\text{NLO},\notag\\
    \Delta X_\text{NNLO} &= \max \left\{\alpha^2X_\text{NLO}, \alpha\left\vert X_\text{NLO} - X_\text{NNLO}\right\vert\right\}.\label{eq:truncation}
\end{align}

\section{Results}

\begin{table}[t]
  \begin{ruledtabular}
    \begin{tabular}{llll}
      & fit & Ref.~\cite{Bijnens:2014lea} & FLAG~\cite{Aoki:2019cca} \\
      \thline
      $(l_2^r - 2l_1^r) \times 10^3$ & $\hphantom{-}12.62(25)(0)$ & $9.9(1.3)$ & $19(17)$ \\
      $l_4^r \times 10^{3}$ & $-2.6(1.1)(0.2)$ & $6.2(1.3)$ & $3.8(2.8)$ \\
    \end{tabular}
  \end{ruledtabular}
    \caption{NLO LECs obtained from a fit to the CLS ensembles (evaluated at $\mu=0.77\,\text{GeV}$).
    The first error is the statistical one, while the second arises due to the error of the lattice spacing.
  For comparison, in the second column the values expected from ChPT analyses are given, while the third contains the values extracted from $N_\text{f} = 2 + 1$ lattice QCD computations~\cite{Aoki:2019cca, Bazavov:2010hj, Beane:2011zm, Borsanyi:2012zv, Durr:2013goa, Boyle:2015exm}.} \label{table:cls_nlo_lecs}
\end{table}  

\begin{table}[t]
  \begin{ruledtabular}
    \begin{tabular}{llll}
      & fit & Ref.~\cite{Bijnens:2014lea} & Refs.~\cite{Bijnens:1997vq,Bijnens:1998fm} \\
      \thline
      $l_1^r \times 10^3$ & $-6.1(1.8)(0.1)$ & $-4.03(63)$ &\\
      $l_2^r \times 10^3$ & $\hphantom{-}2.58(90)(7)$ & $\hphantom{-}1.87(21)$ &\\
      $l_3^r \times 10^3$ & $\hphantom{-}0.776(65)(4)$ & $\hphantom{-}0.8(3.8)$ &\\
      $l_4^r \times 10^3$ & $-33(13)(0)$ & $\hphantom{-}6.2(1.3)$ &\\ 
      $r_a \times 10^6$ & $\hphantom{-}28(12)(1)$ & & $\hphantom{-}13$\\
      $r_b \times 10^6$ & $-4.8(2.6)(0.2)$ & & $-9.0$\\
      $r_c \times 10^6$ & $\hphantom{-}2.1(1.3)(0.1)$ & & $\hphantom{-}1.1$\\
      $r_F^r \times 10^3$ & $\hphantom{-}2.7(1.2)(0)$ & & $\hphantom{-}0$\\
    \end{tabular}
  \end{ruledtabular}
    \caption{The same as Table~\ref{table:cls_nlo_lecs}, but at NNLO.
    For the NNLO LECs we show the estimates from resonance saturation for comparison~\cite{Bijnens:1997vq,Bijnens:1998fm}, although the uncertainties especially in $r_{a,b}$ are substantial and difficult to quantify.}\label{table:cls_nnlo_lecs}
\end{table}

\begin{table}[b]
  \begin{ruledtabular}
    \begin{tabular}{lll}
      & NLO & NNLO \\
      \thline
      $\chi^2/\text{dof}$ & $216 / (122 - 9) = 1.91$ & $165 / (123 - 15) = 1.53$ \\
      BIC & $259$ & $237$ \\
      $M_\rho/\si{MeV}$ & \num[parse-numbers=false]{761.4(5.1)(0.3)(24.7)} & \num[parse-numbers=false]{750(12)(1)(1)} \\
      $\Gamma_\rho/\si{MeV}$ & \num[parse-numbers=false]{150.9(4.4)(0.1)(4.9)} & \num[parse-numbers=false]{129(12)(1)(1)} \\
      $\Re g$ & \num[parse-numbers=false]{5.994(54)(0)(194)} & \num[parse-numbers=false]{5.71(23)(2)(1)} \\
      $-\Im g$ & \num[parse-numbers=false]{0.731(21)(0)(24)} & \num[parse-numbers=false]{0.46(14)(2)(1)} \\
      $ F / \si{\MeV}$ & \num[parse-numbers=false]{88.27(0.23)(0.04)(2.86)} & \num[parse-numbers=false]{93.7(2.3)(0.1)(0.2)} \\
    \end{tabular}
    \caption{Results of NLO and NNLO fits to the CLS data, including 
    the goodness of the fit, the properties of the $\rho$ resonance at the physical point, as well as the decay constant in the chiral limit.
    The first error is the statistical one, the second stems from the lattice spacing, the third is the truncation error estimated via Eq.~\eqref{eq:truncation}. The Bayesian information criterion (BIC) is defined in terms of the number of fit parameters $\vert\mb{F}\vert$ and the number of data points $N$ as $\text{BIC} = \chi^2 + \vert\mb{F}\vert\log N$.}\label{table:cls_pole_positions}
  \end{ruledtabular}
\end{table}

\nocite{Zyla:2020zbs,Kass:1995loi,GarciaMartin:2011jx}

\begin{figure}
    \input{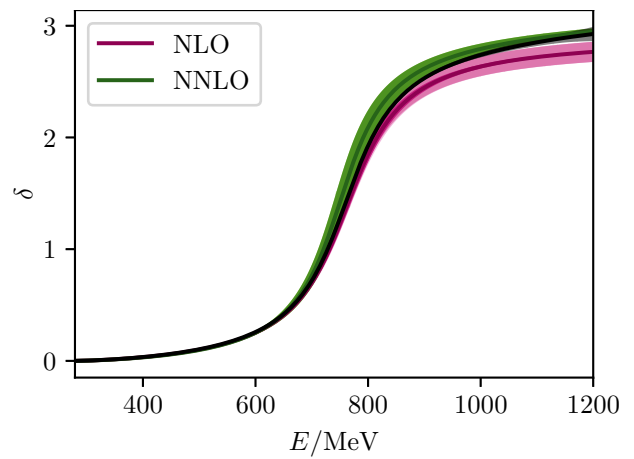}
    \caption{The phase at physical pion mass as extrapolated from global fits to the CLS data;
    see Fig.~\ref{fig:cls_pion_mass_dependence} for color scheme. 
    For comparison, in black the result of the dispersive analysis~\cite{Colangelo:2018mtw}. The extrapolation is performed at fixed $E=\sqrt{s}$, but a trajectory defined by fixed momentum instead would yield identical results.}\label{fig:cls_phase_physical_piont}
\end{figure}

\begin{figure*}
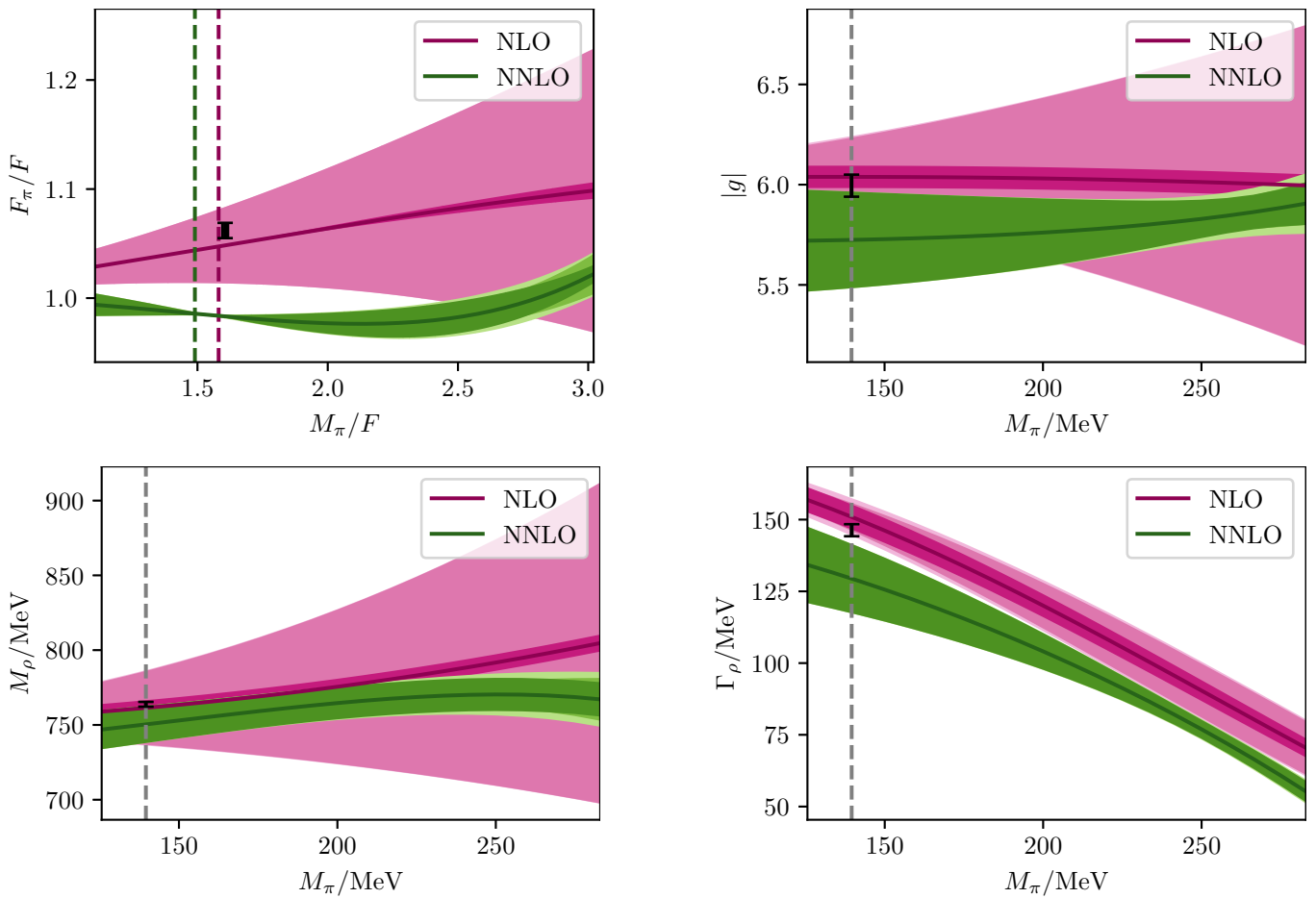

    \subfigure{\input{img/cp3_decay_total.pgf}} \hfill
    \subfigure{\input{img/cp3_coupling_total.pgf}}
    \subfigure{\input{img/cp3_pole_re_total.pgf}} \hfill
    \subfigure{\input{img/cp3_pole_im_total.pgf}}
    \caption{The pion-mass dependence of the decay constant, the coupling, as well as the real and imaginary part of the $\rho$ pole as determined via fits to the CLS data, with error bands corresponding to (in order of decreasing color saturation) the data error (statistical plus spacing), the truncation error, and the total one.
    The dashed lines mark the physical pion mass.
    The decay constant is given in units of $F$ to reduce the impact of the scale setting.
    Since the NLO and NNLO fits yield different values of $F$, their physical points in these units differ.
  Also shown as black ranges are reference values, the $\rho$ characteristics taken from Ref.~\cite{GarciaMartin:2011jx} and the decay constant from Refs.~\cite{Zyla:2020zbs, Aoki:2019cca, Bazavov:2010hj, Beane:2011zm, Borsanyi:2012zv, Durr:2013goa, Boyle:2015exm}.}\label{fig:cls_pion_mass_dependence}
\end{figure*}

To fix the LECs it is necessary to control both the $s$ dependence and the mass dependence.
Hence, we fit all CLS ensembles from Ref.~\cite{Andersen2019} simultaneously, once working to NLO and once working to NNLO, excluding only the ensemble N401 from the fit, since its pion decay constant has not been determined in Ref.~\cite{Bruno:2016plf}.
To render the NNLO fit stable, it is necessary to put a constraint on the LEC $l_3^r$.
This parameter governs the relation between the pion mass $\mpi$ and its value $M$ at LO in ChPT, information on which is not included in our fit. 
Thus we add a penalty term to the $\chi^2$ that favors values of $l_3^r$ around its reference value $0.8(3.8)\times 10^{-3}$~\cite{Bijnens:2014lea}.
The LECs obtained at NLO are given in Table~\ref{table:cls_nlo_lecs}, and the NNLO ones in Table~\ref{table:cls_nnlo_lecs}.
Since the amplitudes as given in Eqs.~\eqref{eq:nlo_iam} and~\eqref{eq:nnlo_iam} have the appropriate analytic structure, they can be continued analytically to the second Riemann sheet, where the pole associated with the $\rho$ resonance is located.
Extracting the mass $M_\rho$ and width $\Gamma_\rho$ from the pole position $s_\text{p}$ via  $s_\text{p} = (M_\rho - i\Gamma_\rho/2)^2$ and the coupling $g$ of $\rho$ to $\pi\pi$ from the residue $r$ via $g^2 = 48\pi r / (4M_\pi^2 - s_\text{p})$ yields the values shown in Table~\ref{table:cls_pole_positions}.
Also shown are the goodness of the fit as well as the obtained value of $F$, the pion decay constant in the chiral limit.
The corresponding phase is depicted in \Figref{fig:cls_phase_physical_piont}.
Here and in the following, the physical point is simply defined by the PDG value of the charged pion mass, $\mpi=\SI{139.57}{\MeV}$~\cite{Zyla:2020zbs}, and $F$ is computed using the PDG value of $F_\pi$ as input.
Because of the unitarization via the IAM,
the LECs are expected to deviate to some extent
from the ChPT reference values~\cite{Dobado1993,Dobado1997,Guerrero:1998ei,GomezNicola:2001as}. Accordingly, all LECs agree well with expectations, apart from a large discrepancy 
in $l_4^r$ both at NLO and NNLO.
To understand its origin, we performed an NLO fit to the pion decay constant alone (at NNLO the fit becomes underconstrained), leading to $l_4^r=1.3(1.0)\times 10^{-3}$, in agreement with FLAG, but already in tension with phenomenology. The remainder of the pull displayed in Table~\ref{table:cls_nlo_lecs} originates from the $\pi\pi$ data. This pull becomes exacerbated at NNLO, but as indicated by the large uncertainties the sensitivity to $l_4^r$ is limited. Indeed, we observe only a moderate increase of the $\chi^2$ if literature values of $l_4^r$ are enforced, as well as a large change to $l_4^r=-16\times 10^{-3}$ when employing a different strategy for the scale setting~\cite{Andersen2019}. We conclude that there is a tension between the pion decay constant in the chiral limit and $\rho$ parameters, which at least in part may be related to scale-setting uncertainties. Further details are given in App.~\ref{sec:appendix_cls_scale_setting}.

In general, we note that the $\chi^2/\text{dof}$ improves significantly when going from NLO to NNLO, although a statistically fully acceptable fit would require a more detailed understanding of lattice artifacts. 
Moreover, in the terms defined in Ref.~\cite{Kass:1995loi}, $\Delta \text{BIC }=22$ provides very strong evidence for the NNLO over the NLO IAM.
Comparing the obtained $\rho$ characteristics with the ones from Roy-like equations~\cite{GarciaMartin:2011jx}---namely $\mrho = 763.7^{+1.7}_{-1.5}\,\si{\MeV}$, $\Gamma_\rho = 146.4^{+2.0}_{-2.2}\,\si{\MeV}$, and $g= 
5.98^{+0.04}_{-0.07}+i (-0.56)^{+0.10}_{-0.07}$---shows that both the NLO and NNLO results are compatible with these already within statistical errors, with a $1.4\sigma$ discrepancy in the width at NNLO and a $2.2\sigma$ tension in $\Im g$ at NLO.
However, only the NLO value of $F$ is compatible with the literature value $F = \SI{86.89+-0.58}{\MeV}$, which is obtained by combining the PDG value of $\Fpi$~\cite{Zyla:2020zbs} with the FLAG $N_\text{f} = 2 + 1$ average of $\Fpi/F$~\cite{Aoki:2019cca, Bazavov:2010hj, Beane:2011zm, Borsanyi:2012zv, Durr:2013goa, Boyle:2015exm}.

Our main results are shown in Figs.~\ref{fig:cls_phase_physical_piont} and~\ref{fig:cls_pion_mass_dependence}, for the pion-mass dependence of the phase shift, the decay constant, and the $\rho$ resonance parameters. Most notably, the two-loop analysis allows us to improve the precision considerably when going beyond the physical point, once the truncation becomes the dominant source of error.
Second, with error bands produced assuming a breakdown scale of $\mrho$, the NLO and NNLO bands mostly overlap, which indicates that the true breakdown scale of the theory may lie below the $\rho$ mass, but not by much. 

Overall, the coupling shows a very mild mass dependence~\cite{Hanhart:2008mx,Pelaez:2010fj}, as does the $\rho$ mass. 
Towards the end of the fit range, the central value of the two-loop curve seems to decrease, in disagreement with the phenomenological expectation from both the KSFR relation~\cite{Kawarabayashi:1966kd,Riazuddin:1966sw} and the expected ordinary $q\bar q$ nature of the $\rho$ meson~\cite{RuizdeElvira:2017aet}. 
This may again, in addition to the $\chi^2$ and the tension in $l_4^r$, point to the impact of lattice artifacts, which the two-loop IAM becomes flexible enough to mimic.   

Similar conclusions can be drawn from the fits to the data by HadSpec, which are described in App.~\ref{sec:appendix_HadSpec_results}.
They also show a significant improvement of the $\chi^2$ when going from NLO to NNLO and a pion-mass dependence that mimics the one depicted in Fig.~\ref{fig:cls_pion_mass_dependence}, with the difference that $M_\rho$ does not decrease at high pion masses, providing further evidence that this decrease may arise due to lattice artifacts.
Notably, for the HadSpec data the $\rho$ properties at the physical point are closer to the literature values at NLO than at NNLO.

\section{Conclusions}
In this work we have presented compact analytic expressions for the two-loop partial-wave amplitudes for $\pi\pi$ scattering up to $D$-waves, with a first application to an analysis of lattice data for the $P$-wave amplitude and the $\rho$ parameters. We have shown that two-loop fits do improve the fit quality and, by comparing NLO and NNLO results, found that the breakdown scale of the chiral expansion should not lie much below the expected scale set by the $\rho$ mass. However, we also concluded that the current data sets cannot be described in a statistically satisfactory way, with a more detailed understanding of the lattice data required. 

In the future, anticipated improvements in the precision of lattice QCD calculations will increase the need to match that precision in the analysis. In this work, we have demonstrated how to achieve two-loop precision in practice, using the example of the $P$-wave, but once lattice calculations mature a similar analysis can be performed for other partial waves including the pion-mass dependence of the $f_0(500)$. Even once data sets at the physical point become available, the IAM will thus provide a tool for a high-precision analysis of lattice data.

\begin{acknowledgments}
  Finite-volume energy levels taken from Refs.~\cite{Dudek2013,Wilson2015} were provided by the Hadron Spectrum Collaboration (HadSpec)---no endorsement on their part of the analysis presented in the current paper should be assumed.
  In addition, we would like to thank John Bulava as well as ETMC for sharing lattice data with us.
We thank Carsten Urbach for many useful discussions, and Ra\'ul Brice\~no, Mattia Bruno, Christopher Thomas, Martin Ueding, Markus Werner, and David Wilson for valuable input.
Financial support by the Bonn--Cologne Graduate School of Physics and Astronomy (BCGS), the DFG (CRC 110, ``Symmetries and the Emergence of Structure in QCD''), the Swiss National Science Foundation, under Project Nos.\ PCEFP2\_181117 (M.H.) and PZ00P2\_174228 (J.R.E.), and the DOE (Grant No.\ DE-FG02-00ER41132) is gratefully acknowledged.                          
\end{acknowledgments}

\appendix

\begin{widetext}  
  \section{Details of lattice data and fit strategy}\label{sec:appendix_lattice_details}
To understand the precise definition of the $\chi^2$, first it is expedient to briefly recall L\"{u}scher's method in the context of the $P$-wave of interest.
The procedure is roughly as follows~\cite{Briceo2018}:
on the lattice, operators corresponding to discrete-world versions of $\pi\pi$ states with $I=J=1$ are constructed.
These states are characterized by their irreducible representation (irrep) of the residual rotational symmetry as well as their relative momentum with respect to the rest frame, i.e., the boost momentum $\vec{d}\in\mb{Z}^3$.
Subsequently, the energy levels of these states are computed.
Such an energy level $E^\text{lat}$ is related to the scattering phase shift $\delta$ via the quantization condition
\begin{equation}\label{eq:quantization_condition}
  \delta\left(E^\text{lat}\right) = \mc{Z}\left(E^\text{lat}\right),
\end{equation}
where $\mc{Z}$ is a known, albeit complicated, expression that can be computed only numerically and depends on the details of the lattice, the pion mass, as well as the irrep and the boost of the state with energy $E^\text{lat}$.
A given data set (ensemble) contains several energy levels $E^\text{lat}_i$, $i = 1,\dots,N$, corresponding to different combinations of boost and irrep as well as different excitations of the states.
Moreover, it has a fixed pion mass and, if it has been measured, pion decay constant.
Both the CLS collaboration and HadSpec have generated several data sets with different characteristics, the ones of relevance for this work are listed in Table~\ref{table:ensembles}.
\begin{table}[t]
  \begin{ruledtabular}
    \begin{tabular}{cccc}
      \multicolumn{2}{c}{CLS} & \multicolumn{2}{c}{HadSpec} \\
      \thline
      name & $\mpi / \si{\MeV}$ & name & $\mpi / \si{\MeV}$ \\
      \thline
      N401 & 284 & 20 & 391 \\
      N200 & 283 & 24 & 391 \\
      J303 & 258 & 32 & 236 \\
      C101 & 223 \\
      D101 & 223 \\
      D200 & 200 \\
    \end{tabular}
  \end{ruledtabular}
  \caption{The lattice ensembles with approximate values of the pion mass.}\label{table:ensembles}
\end{table}
For simplicity, consider first a single ensemble in isolation.
To fit the IAM to the data of this ensemble, we proceed as follows:
for a fixed choice of LECs, irrep, and boost the phase $\delta^\text{IAM}(E) = \arg(t(E))$, with $t$ given by either Eq.~\eqref{eq:nlo_iam} or Eq.~\eqref{eq:nnlo_iam}, is inserted into \Eqref{eq:quantization_condition}, which is subsequently solved numerically to determine an energy $E^\text{IAM}$, corresponding to the fixed boost and irrep.
This is done for all combinations of boost, irrep, and all excitations, yielding one energy level $E^\text{IAM}_i$ for each level $E^\text{lat}_i$.
The goal is then to minimize the distance of these energies via a variation of the LECs.
More specifically, we introduce a $\chi^2$ of the form $\chi_\text{single}^2 = v^TC^{-1}v$, $v = v^\text{lat} - v^\text{model}$, with
\begin{equation}
    v^\text{lat} =
    \begin{pmatrix}
        F_\pi^\text{lat} \\ \mpi^\text{lat} \\ E^\text{lat}_1 \\ E^\text{lat}_2 \\ \vdots \\ E^\text{lat}_N
    \end{pmatrix}, \qquad
    v^\text{model} =
    \begin{pmatrix}
       F_\pi^\text{ChPT}\left(F, \mpi; \mb{L}_1\right) \\ \mpi \\ E^\text{IAM}_1\left(F, \mpi; \mb{L}_2\right) \\ E^\text{IAM}_2\left(F, \mpi; \mb{L}_2\right) \\ \vdots \\ E^\text{IAM}_N\left(F, \mpi; \mb{L}_2\right)
    \end{pmatrix},
\end{equation}
and $C$ the covariance matrix of the lattice data.
Here $F_\pi^\text{ChPT}$ is given by the expressions in App.~\ref{sec:appendix_pion_decay_chpt}, $\mb{L}_1$ and $\mb{L}_2$ are the set of LECs on which the model expressions for the decay constant and the phase shift depend, respectively, and $\mpi$ as well as $F$ are additional fit parameters. 
That is, the fit parameters for a single ensemble are $\mb{F} = \{F\} \cup \{\mpi\} \cup \mb{L}_1 \cup \mb{L}_2$.
$M_\pi$ is introduced as a fit parameter to take into account the error of the pion mass, while treating $F$ as a fit parameter allows us to work almost exclusively in lattice units, see App.~\ref{sec:appendix_error}.
We always fit multiple ensembles simultaneously, because in this way the pion-mass dependence of both the IAM and $F_\pi^\text{ChPT}$ can be controlled, which becomes particularly important at NNLO, where several free parameters are present.
Since the data on different ensembles are not correlated, the  $\chi^2$ is given as $\chi^2 = \sum_{k} \chi^2_k$, with each $\chi^2_k$ mimicking $\chi^2_\text{single}$.
If two ensembles share the same pion mass or decay constant, the corresponding entry is taken into account only once.
Moreover, since $F$ in physical units is the same for all ensembles, there is only one fit parameter $F$ for each lattice spacing.
The minimization of the $\chi^2$ requires repeated evaluations of the quantity $\mc{Z}$ in \Eqref{eq:quantization_condition}, whose computation is numerically demanding.
To accelerate it, we slightly modify the approach of Ref.~\cite{Hu:2017wli}:
assuming that the fit is not completely off, for each energy level $E^\text{lat}$ it will yield an energy level $E^\text{IAM}$ that is close to the former, i.e., $(E^\text{lat} - E^\text{IAM}) / E^\text{lat} \ll 1$.
The same holds true for the mass.
Hence it is justified to Taylor-expand both sides of the quantization condition $\delta^\text{IAM}(E^\text{IAM}) = \mc{Z}(E^\text{IAM})$ around $E^\text{lat}$ and $M_\pi^\text{lat}$, up to and including linear order.
Solving the result for $E^\text{IAM}$ yields
\begin{equation}\label{eq:quantization_taylor}
  E^\text{IAM}
  = E^\text{lat} - \frac{\big[\mc{Z}-\delta^\text{IAM}\big]\left(E^\text{lat}, \mpi^\text{lat}\right)
  + \left(\mpi-\mpi^\text{lat}\right)\Big[\frac{\partial(\mc{Z}-\delta^\text{IAM})}{\partial M}\Big]\left(E^\text{lat}, \mpi^\text{lat}\right)}{\Big[\frac{\partial(\mc{Z}-\delta^\text{IAM})}{\partial E}\Big]\left(E^\text{lat}, \mpi^\text{lat}\right)}.
\end{equation}
Thus, for a given energy $E^\text{lat}$ and mass $\mpi^\text{lat}$ we need to compute $\mc{Z}$ and its first derivatives only once, we do so by numerically differentiating the full quantization condition~\eqref{eq:quantization_condition}.
Then we can compute $E^\text{IAM}$ for different values of the LECs by merely re-evaluating the IAM.
This speeds up the minimization drastically.
We explicitly checked that this approximation is justified by performing selected fits twice, once using the full quantization condition and once using Eq.~\eqref{eq:quantization_taylor}, both yielding the same minimum, i.e., the same values of the LECs.
However, the value of the $\chi^2$ obtained using the Taylor expansion tends to be slightly (at most a few percent) bigger than the exact value, hence to obtain the correct $\chi^2$ all results are re-evaluated using the exact quantization condition.
Note that this deviation in the $\chi^2$ is expected:
the $\chi^2$ depends on the LECs only via the IAM (ignoring $F_\pi^\text{ChPT}$ for the time being, since it is of no relevance for this argument), so we might write $\chi^2(\{E^\text{IAM}_i(\mb{L}_2)\})$.
A successful minimization will result in values $\mb{L}_2^\text{min}$ for which $\chi^2$ is at a global minimum.
Equation~\eqref{eq:quantization_taylor} then amounts to an approximation of the energy values for which $\chi^2$ is minimal, therefore its value needs to increase.
Even with this acceleration at hand, the minimization problem provides a challenge.
Namely at NNLO, there are eight LECs.
In addition, there are the fit parameters $\mpi$ and $F$, e.g., in a global fit to the CLS data four fit parameters corresponding to different pion masses and three fit parameters $F$ corresponding to different lattice spacings are required.
Altogether, there are up to 15 fit parameters (even more if the error of the lattice spacing is taken into account, see App.~\ref{sec:appendix_error}).
To still obtain reliable results, a powerful minimization algorithm is needed.
The differential evolution algorithm~\cite{Storn1997} turns out to be a good choice, i.e., it allows for stable fits while still consuming not too much time.
Roughly, it works via generating a population of trial solutions, recombining them randomly into a new generation to find solutions of a better fitness (in our case a lower value of the $\chi^2$), and iterate this.
To make it work properly, it needs to be configured carefully, a procedure that requires repeated iterations of the same fits while varying the configuration parameters and observing the impact on stability, the resulting $\chi^2$, and performance.
In addition to the population size, there are two core handles:
the differential weight and the crossover probability, influencing how strongly the components and how many components of the trial solutions are modified in each iteration, respectively.
For the problem at hand, population sizes of $20\vert\mb{F}\vert$ for fits to the CLS data and $40\vert\mb{F}\vert$ for fits to the data by HadSpec are sufficient, where $\mb{F}$ is the set of all fit parameters.
With these population sizes, the crossover probability can be set to 1, while the differential weight is randomly changed to a value between 0.5 and 1 before each iteration, a technique known as dithering.
The results obtained in this way are refined via Powell's method~\cite{Press:2007num}; for both algorithms we use the implementation in \texttt{SciPy}~\cite{2020SciPy-NMeth}.
It comes at no surprise that the data by HadSpec, which have a lower number of data points and pion masses as compared to the CLS data, require more generous settings of the differential evolution algorithm, i.e., a larger population size relative to the number of fit parameters.
Exemplarily, the energy levels as well as the phase as obtained in the NNLO fit to the CLS data are shown in \Figref{fig:d101_data_vs_nnlo_fit} for a single ensemble.

\begin{figure}[t]
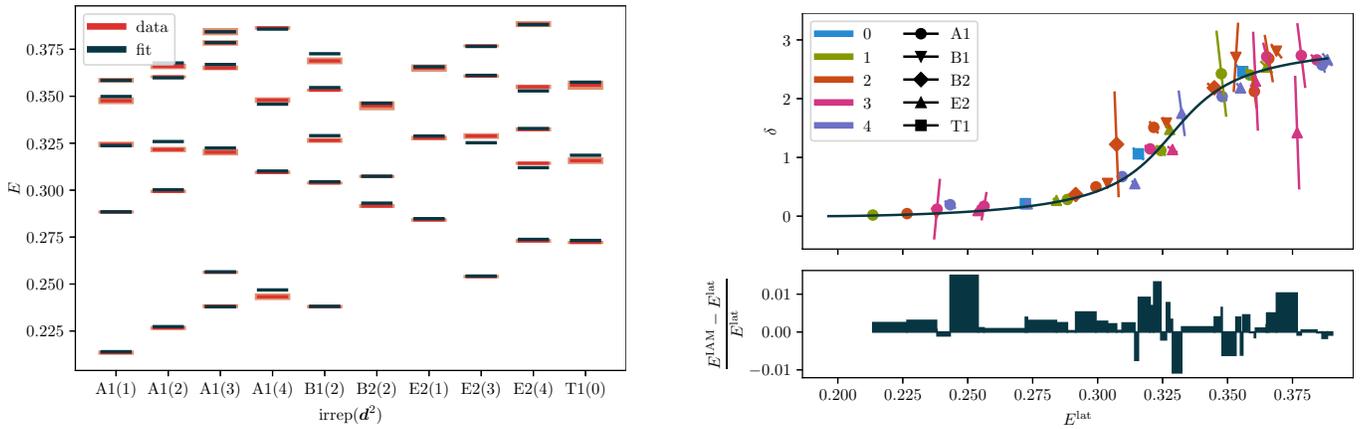

      \subfigure{\scalebox{0.62}{\input{img/223_D101_spectrum.pgf}}} \hfill
      \subfigure{\scalebox{0.62}{\input{img/223_D101_phase.pgf}}}
    \caption{Results of a global NNLO fit to the CLS data compared to the data of the ensemble D101.
      (Left) The energy levels, each column corresponds to a different combination of boost and irrep, e.g., the leftmost column contains the energy levels of states in the irrep A1, with $\vec{d}^2=1$ the value of the squared boost momentum.
    (Right) The corresponding scattering phase.
    Here the color encodes $\vec{d}^2$, while the shape encodes the irrep.
    The curve corresponds to the fit result.
    Also shown are the relative differences of the lattice energies and the ones obtained in the fit.
    }\label{fig:d101_data_vs_nnlo_fit}
\end{figure}

\section{Error computation and impact of lattice spacing}\label{sec:appendix_error}
Taking into account the statistical error of the data is straightforward:
both CLS and HadSpec provide each energy level $E^\text{lat}$ as a collection of several hundred values, each corresponding to a different sample of the underlying gauge configurations.
Via jackknife resampling of these underlying values and repeating the fit on each jackknife sample the error can be computed taking into account the correlation of the energy levels automatically.
The reason to pick a jackknife (i.e., drawing samples via omitting a single underlying value in each run) instead of a bootstrap (i.e., sampling the underlying values randomly with replacement) lies in the nature of the quantization condition~\eqref{eq:quantization_condition}~\cite{Werner:2019hxc}.
It has poles at energies that correspond to free $\pi\pi$ states.
Sometimes, the values $E^\text{lat}$ are so close to these poles that resampling the underlying value via a bootstrap yields central values on the other side of the nearby pole, thereby resulting in a completely different value of the phase.
The use of jackknife resampling circumvents this issue, for jackknife samples are significantly more narrowly distributed than bootstrap samples.
There is one subtlety stemming from the jackknife.
It needs the same number of underlying data points for all energy levels.
However, the HadSpec ensembles differ in this respect.
Accordingly, in a global fit to the HadSpec data we feed the jackknife errors into a parametric bootstrap, as suggested in Ref.~\cite{Werner:2019hxc}.
The other source of error associated with the data is the scale-setting.
Namely, on the lattice all quantities are computed in units of the lattice spacing $a$, the distance between two adjacent sites in one direction.
To translate the quantities into physical units, they need to be multiplied by the appropriate power of $a$.
Thus $a$ needs to be determined in physical units, and this so called scale-setting is error-prone.
For example, in Ref.~\cite{Bruno:2016plf} two different methods are used to set the scale of the CLS ensembles, strategy 1 (via the Wilson flow) and strategy 2 (via decay constants), both yielding different results that are incompatible within their errors.
Hence throughout the entire fit we work in lattice units wherever possible.
However, there is one place where the lattice spacing enters the fit: the renormalization scale $\mu$ appearing in the ChPT expressions for both the scattering amplitude and the pion decay constant, see App.~\ref{sec:ChPT}.
It shows up both explicitly via logarithms and implicitly, since the LECs depend on $\mu$ in a way that renders the total amplitude scale-independent.
Therefore, to have one single set of LECs that can be used on all ensembles in a global fit and for extrapolations of observables to the physical point, $\mu$ needs to be fixed globally, and we make the common choice $\mu=\SI{770}{MeV}$.
When fitting, this requires the translation of $\mu$ into lattice units via $\mu\mapsto a\mu$ inside the logarithms.
This is the only place where the lattice spacing shows up in our fits.
To compute the impact of the error of $a$ (with a fixed choice of scale-setting) on the fit results, we extend the $\chi^2$:
\begin{equation}
  \chi^2\mapsto
  \chi^2\Big\vert_{\vec{a}^\text{lat}\mu\mapsto \vec{a}^\text{fit}\mu}
  + \sum_{ij}\left(a^\text{lat}_i - a^\text{fit}_i\right)\left[C_a^{-1}\right]_{ij}\left(a^\text{lat}_j - a^\text{fit}_j\right).
\end{equation}
That is, for each lattice spacing $a_i^\text{lat}$ we introduce a fit parameter $a^\text{fit}_i$, with $C_a$ the covariance matrix of the lattice spacings.
The fit is then repeated multiple times with different values of the lattice spacings, obtained via drawing samples from a multivariate normal distribution with means $a_i^\text{lat}$ and covariance matrix $C_a$.
To estimate $C_a$, we proceed differently depending on the data set.
The HadSpec data are obtained at two different lattice spacings, corresponding to the two different pion masses in Table~\ref{table:ensembles}.
To set the scale, the mass of the $\Omega$ baryon as determined on the lattice~\cite{Wilson2015,Edwards:2011jj} is divided by its experimental value, i.e., $a^\text{lat}_i = M_{\Omega, i}^\text{lat} / M_\Omega^\text{exp}$, $i = 1, 2$.
The two spacings are correlated due to the common choice of $M_\Omega^\text{exp}$.
We take this into account by resampling $M_\Omega^\text{exp}$ using its PDG value and error~\cite{Zyla:2020zbs}, for each value obtained in this way we sample $M_{\Omega, i}^\text{lat}$ and compute $a_i^\text{lat}$, $i=1,2$.
All samples are drawn parametrically assuming a normal distribution.
Repeating this multiple times allows us to estimate $C_a$ in the standard way.
Given the small error of $M_\Omega^\text{exp}$, the off-diagonal entries of $C_a$ are an order of magnitude smaller than the diagonal ones.
The CLS data are obtained at three different lattice spacings (ignoring the ensemble N401, which is excluded from the fits since the corresponding pion decay constant is not available).
The procedure in case of the scale-setting via strategy 1 is very similar to the one outlined for the HadSpec ensembles, with the quantity $t_0$ associated with the Wilson flow replacing the $\Omega$ mass and all relevant values given in Ref.~\cite{Bruno:2016plf}.
However, since the error of the reference value of $t_0$ is larger than the one of $M_\Omega^\text{exp}$, in this case the off-diagonal entries of $C_a$ are sizable.
On the other hand, $C_a$ for strategy 2 is assumed to be diagonal.
The results described in the main text were obtained using strategy 1, since it is the one preferred by the authors of Ref.~\cite{Bruno:2016plf}.
Since the lattice spacings show up only via the renormalization scale in logarithms, the impact of their errors is expected to be small.
This is indeed what is found, as becomes clear in Tables~\ref{table:cls_nlo_lecs}, \ref{table:cls_nnlo_lecs}, \ref{table:cls_pole_positions}: the statistical error always dominates the one stemming from the lattice spacing.
Note also that the latter increases when going from NLO to NNLO, the reason being the more dominant role the logarithms containing $\mu$ play at NNLO, where they appear not only in $F_\pi$, but also in the scattering amplitude.

\section{Scale setting in CLS fits}\label{sec:appendix_cls_scale_setting}
As observed in the main text, fitting the CLS data using strategy 1 for the scale-setting (see App.~\ref{sec:appendix_error}) yields values of $l_4^r$ both at NLO and NNLO that are in conflict with literature values.
To understand this discrepancy, a detail of the scale-setting deserves further attention:
as explained in Ref.~\cite{Bruno:2016plf}, strategy 1 requires small shifts in the values of the pion masses and decay constants as measured on the CLS ensembles (since the statistical uncertainties in strategy 2 are significantly larger than in strategy 1, these required mass corrections are neglected in strategy 2).
However, the $\pi\pi$ energy levels from Ref.~\cite{Andersen2019} are obtained using non-shifted pion masses.
Thus, to be consistent, we also use the non-shifted values of the decay constants and masses in the fit, at the price of being inconsistent with strategy 1.
To assess if this inconsistency is the reason for the discrepancy in $l_4^r$, we re-perform both the NLO and NNLO fits, this time using the lattice spacings of strategy 2.
In Tables~\ref{table:cls_nlo_strat1_vs_2} and~\ref{table:cls_nnlo_strat1_vs_2}, the resulting LECs are compared with the ones obtained previously.
Clearly, both at NLO and NNLO the central value of $l_4^r$ moves a little closer to its literature values, but the discrepancy remains sizable.
%

\begin{table}
  \begin{ruledtabular}
    \begin{tabular}{lllll}
      & strategy 2 & strategy 1 & Ref.~\cite{Bijnens:2014lea} & FLAG~\cite{Aoki:2019cca} \\
      \thline
      $(l_2^r - 2l_1^r) \times 10^3$ & $\hphantom{-}12.64(25)(1)$ & $\hphantom{-}12.62(25)(0)$ & $9.9(1.3)$ & $19(17)$ \\
      $l_4^r \times 10^{3}$ & $-2.0(1.1)(0.2)$ & $-2.6(1.1)(0.2)$ & $6.2(1.3)$ & $3.8(2.8)$ \\
    \end{tabular}
  \end{ruledtabular}
  \caption{NLO LECs obtained from a fit to the CLS ensembles, using different strategies to set the scale.}\label{table:cls_nlo_strat1_vs_2}
\end{table}

\begin{table}
  \begin{ruledtabular}
    \begin{tabular}{lllll}
      & strategy 2 & strategy 1 & Ref.~\cite{Bijnens:2014lea} & Refs.~\cite{Bijnens:1997vq,Bijnens:1998fm} \\
      \thline
      $l_1^r \times 10^3$ & $-2.5$ & $-6.1(1.8)(0.1)$ & $-4.03(63)$ &\\
      $l_2^r \times 10^3$ & $\hphantom{-}1.48$ & $\hphantom{-}2.58(90)(7)$ & $\hphantom{-}1.87(21)$ &\\
      $l_3^r \times 10^3$ & $\hphantom{-}0.822$ & $\hphantom{-}0.776(65)(4)$ & $\hphantom{-}0.8(3.8)$ &\\
      $l_4^r \times 10^3$ & $-16$ & $-33(13)(0)$ & $\hphantom{-}6.2(1.3)$ &\\ 
      $r_a \times 10^6$ & $\hphantom{-}17$ & $\hphantom{-}28(12)(1)$ & & $\hphantom{-}13$\\
      $r_b \times 10^6$ & $-1.7$ & $-4.8(2.6)(0.2)$ & & $-9.0$\\
      $r_c \times 10^6$ & $\hphantom{-}0.24$ & $\hphantom{-}2.1(1.3)(0.1)$ & & $\hphantom{-}1.1$\\
      $r_F^r \times 10^3$ &$\hphantom{-}1.4$ & $\hphantom{-}2.7(1.2)(0)$ & & $\hphantom{-}0$\\
    \end{tabular}
  \end{ruledtabular}
    \caption{Same as Table~\ref{table:cls_nlo_strat1_vs_2}, but at NNLO.}\label{table:cls_nnlo_strat1_vs_2}
\end{table}

%
To check if the discrepancy can be further reduced, we again perform the strategy 2 fits, this time putting a constraint on $l_4^r$ to enforce it to be close to its literature value $6.2(1.3)\times 10^{-3}$~\cite{Bijnens:2014lea}.
At NLO, this leads to an increase of the $\chi^2/\text{dof}$, with the resulting value of $l_4^r = 0.4\times 10^{-3}$, still not perfectly overlapping with the reference value.
An increase of the $\chi^2/\text{dof}$ can also be observed at NNLO, with the resulting value of $l_4^r = 5.4\times 10^{-3}$ compatible with the reference value.
However, the values of the other LECs change considerably, most notably $l_2^r = -0.21\times 10^{-3}$ now being in conflict with the literature, a shift that can already be anticipated by having a look at the correlation $c=\SI{-98}{\percent}$ of $l_2^r$ and $l_4^r$ in the strategy 1 fit.
In addition, the value of $F$ improves, namely its central value moves from $\SI{90.8}{\MeV}$ to $\SI{87.2}{\MeV}$.
On the other hand, the $\rho$ characteristics at the physical point get worse, the pole is now located at $\sqrt{s_\text{p}} = (816 - 78i)\,\si{\MeV}$ instead of $\sqrt{s_\text{p}} = (756 - 64i)\,\si{\MeV}$.
To pin down the source of this behavior, we perform fits to the decay constants only as a function of the pion mass.
This forces us to set the scale directly in the beginning, since otherwise the fit would be underconstrained.
Even with the scale set, an NNLO fit is impossible, for at this order there are six free parameters.
Hence we work at NLO, with $F$ and $l_4^r$ as the sole free parameters.
The results are given in Table~\ref{table:cls_decay_only} and depicted in Fig.~\ref{fig:cls_decay_only}.
Within the statistical error originating in $F_\pi^\text{lat}$ and $\mpi^\text{lat}$, strategy 2 is compatible with the literature values of the two LECs, although it is in tension with the one from phenomenology.
However, the fit using strategy 1 yields values that are incompatible with the literature, shifting the data does not improve the situation significantly.

\begin{figure}[t]
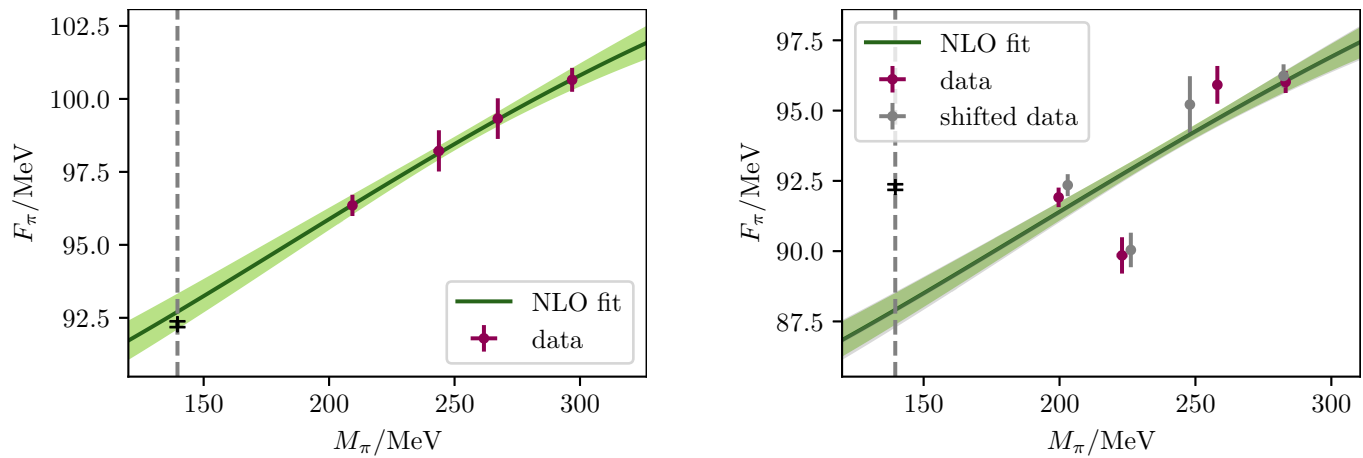

  \subfigure{\input{img/cls_decay_only_strategy_2.pgf}} \hfill
 \subfigure{\input{img/cls_decay_only_strategy_1.pgf}}
 \caption{The pion decay constant as determined in a fit to the decay constants of the CLS ensembles.
 The error bands show the statistical error. Also shown as a gray dotted line is the physical point, in addition, in black the PDG value~\cite{Zyla:2020zbs}.
 (Left) The fit using strategy 2 to set the scale.
 (Right) The fit with strategy 1. The shifted data are shown in gray, as is the corresponding fit, which is almost indistinguishable from the one to the non-shifted data, shown in green.}\label{fig:cls_decay_only}
\end{figure}

\begin{table}
  \begin{ruledtabular}
    \begin{tabular}{lllll}
      & strategy 2 & strategy 1 & strategy 1 (shifted) & reference \\
      \thline
      $F/\si{MeV}$ & $87.72(87)$ & $82.51(84)$ & $82.48(93)$ & $86.89(58)$ \\
      $l_4^r\times 10^3$ & $0.8(1.0)$ & $1.3(1.0)$ & $1.3(1.1)$ & $3.8(2.8)$, $6.2(1.3)$\\
    \end{tabular}
  \end{ruledtabular}
  \caption{The LECs obtained in fits to the pion decay constants on the CLS ensembles.
  The error is the statistical one, stemming both from $F_\pi^\text{lat}$ and $\mpi^\text{lat}$.
  The reference value of $F$ and the first one given for $l_4^r$ are taken from Refs.~\cite{Zyla:2020zbs, Aoki:2019cca, Bazavov:2010hj, Beane:2011zm, Borsanyi:2012zv, Durr:2013goa, Boyle:2015exm}, the second reference value of $l_4^r$ from Ref.~\cite{Bijnens:2014lea}.}\label{table:cls_decay_only}
\end{table}

These observations point to the following conclusion:
independently of the scale setting strategy, the full IAM fits to the CLS $\pi\pi$ data require values of $F$ that are larger than the literature value.
This, in turn, pushes the values of $l_4^r$ to unphysically small values.
Since a fit to the decay constants only using strategy 2 produces values that are more in agreement with the literature, the discrepancy seems to emanate from the $\rho$ data.
However, the shift in $l_4^r$ when going from strategy 1 to strategy 2 indicates that scale-setting effects do play an important role.
We stress that this analysis is not the final word, an improved fit to the decay constants only (taking into account the error of the lattice spacing) and a better understanding of the overall lattice artifacts would be necessary to completely settle this issue. Ultimately, these observations are likely related to the fact that even at two-loop order the quality of the resulting fits to the lattice data cannot be considered statistically acceptable.  

\section{Fits to HadSpec data}\label{sec:appendix_HadSpec_results}
Compared to the CLS data, the data by HadSpec have fewer energy levels at only two different pion masses, see Table~\ref{table:ensembles}.
Nevertheless, stable fits are possible if more generous settings of the minimization algorithm are used, as explained in App.~\ref{sec:appendix_lattice_details}.
On top of this, it is necessary to set $l_3^r$ to a fixed value, a constraint alone is insufficient, for the fit is insensitive to the literature value within its errors.
Hence we fix $l_3^r = 8.2\times 10^{-4}$.
Another crucial difference is that the pion decay constant has not been measured on the ensemble with the lower pion mass.
Contrary to what has been done in the case of the N401 CLS ensemble, excluding the ensemble from the fit is not an option, because this would leave only a single pion mass.
Consequently, instead of fitting the pion decay constant, we set $F$ to its literature value $F = \SI{86.89+-0.58}{\MeV}$~\cite{Zyla:2020zbs, Aoki:2019cca, Bazavov:2010hj, Beane:2011zm, Borsanyi:2012zv, Durr:2013goa, Boyle:2015exm}.
Since $F$ is needed in lattice units, the lattice spacing appears now both in the translation $\mu\mapsto a\mu$ and $F\mapsto aF$, increasing its relevance for an error computation.
In addition, the error of the literature value itself needs to be taken into account.
The numerical results of a global fit at NLO and NNLO are shown in Tables~\ref{table:HadSpec_nlo_lecs},~\ref{table:HadSpec_nnlo_lecs}, and~\ref{table:HadSpec_pole_positions}, while the phase at the physical point as well as the pion-mass dependence of the $\rho$ characteristics are depicted in Fig.~\ref{fig:HadSpec_results}.
The conclusions drawn from the fits to the CLS ensembles carry over in large parts, with a few differences:
due to the more important role of the lattice spacing, its error is sometimes dominant.
In addition, the pole position of the $\rho$ at the physical point at NNLO deviates significantly from its literature value, with both the mass and width being too low.
On the other hand, the mass does not start to decrease at high pion masses, as opposed to the CLS fit and in agreement with phenomenology~\cite{Kawarabayashi:1966kd,Riazuddin:1966sw, RuizdeElvira:2017aet}.

\begin{figure}
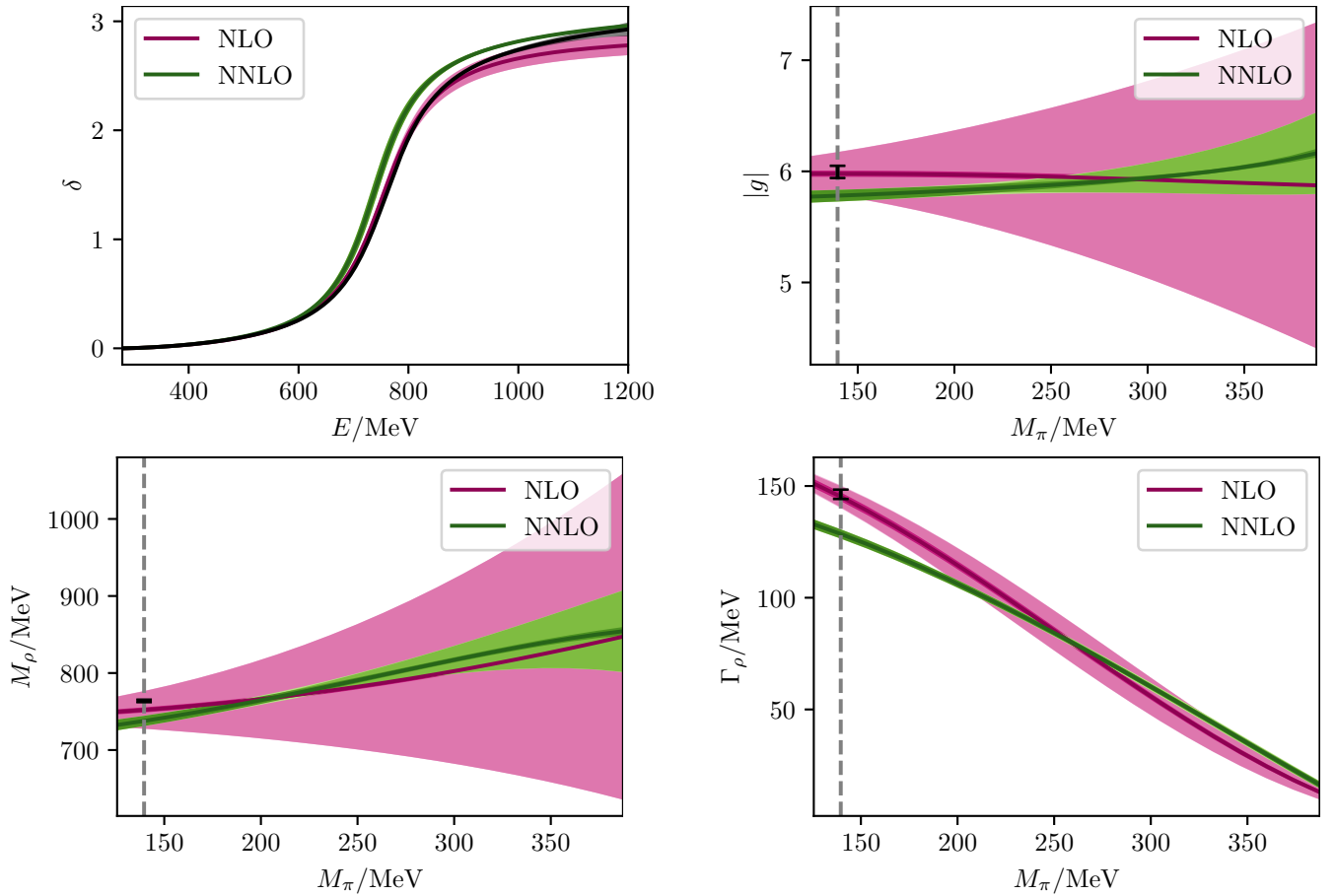

\scalebox{0.99}{
 \hspace{0.5cm} \subfigure{\input{img/hadspec_phase_total.pgf}} \hspace{1.0cm} 
  \subfigure{\input{img/hadspec_coupling_total.pgf}}}
  \scalebox{0.99}{
  \subfigure{\input{img/hadspec_pole_re_total.pgf}} \hspace{1.1cm}
  \subfigure{\input{img/hadspec_pole_im_total.pgf}}}
  \caption{The phase at the physical point, as well as the pion-mass dependence of the coupling, the real part, and imaginary part of the $\rho$ pole as determined via fits to the HadSpec data.
  The error bands are as in Fig.~\ref{fig:cls_pion_mass_dependence}, with the data error now including the error of the literature value of $F$.}\label{fig:HadSpec_results}
\end{figure}

\begin{table}
  \begin{ruledtabular}
    \begin{tabular}{llll}
      & fit & Ref.~\cite{Bijnens:2014lea} & FLAG \\
      \thline
      $(l_2^r - 2l_1^r) \times 10^3$ & $\hphantom{-}12.908(59)(68)(99)$ & $9.9(1.3)$ & $19(17)$ \\
    \end{tabular}
  \end{ruledtabular}
    \caption{NLO LEC obtained from a fit to the HadSpec ensembles.
    The first error is the statistical one, the second arises due to the error of the lattice spacing, the third stems from the error of the literature value of $F$.
  For comparison, in the second column the values expected from ChPT analyses are given, while the third contains the values extracted from $N_\text{f} = 2 + 1$ lattice QCD computations~\cite{Aoki:2019cca, Bazavov:2010hj, Beane:2011zm, Borsanyi:2012zv, Durr:2013goa, Boyle:2015exm}.}\label{table:HadSpec_nlo_lecs}
\end{table}

\begin{table}
  \begin{ruledtabular}
    \begin{tabular}{llll}
      & fit & Ref.~\cite{Bijnens:2014lea} & Refs.~\cite{Bijnens:1997vq,Bijnens:1998fm} \\
      \thline
      $l_1^r \times 10^3$ & $-4.17(85)(24)(32)$ & $-4.03(63)$ \\
      $l_2^r \times 10^3$ & $\hphantom{-}1.44(25)(13)(18)$ & $\hphantom{-}1.87(21)$ \\
      $r_a \times 10^6$ & $\hphantom{-}11.5(2.0)(0.4)(0.6)$ & & $\hphantom{-}13$\\
      $r_b \times 10^6$ & $-4.39(84)(25)(25)$ & & $-9.0$\\
      $r_c \times 10^6$ & $\hphantom{-}0.97(0.43)(0.12)(0.16)$ & & $\hphantom{-}1.1$\\
    \end{tabular}
  \end{ruledtabular}
  \caption{Same as Table~\ref{table:HadSpec_nlo_lecs}, but at NNLO, more details on the reference values can be found in Table~\ref{table:cls_nnlo_lecs}.}\label{table:HadSpec_nnlo_lecs}
\end{table}

\begin{table}
  \begin{ruledtabular}
    \begin{tabular}{lll}
      & NLO & NNLO \\
      \thline
      $\chi^2/\text{dof}$ & $124 / (52 - 4) = 2.58$ & $75 / (52 - 8) = 1.70$ \\
      BIC & $140$ & $106$ \\
      $M_\rho/\si{MeV}$ & \num[parse-numbers=false]{752.1(2.0)(2.4)(1.2)(24.4)} & \num[parse-numbers=false]{737.8(3.8)(5.4)(0.6)(0.8)} \\
      $\Gamma_\rho/\si{MeV}$ & \num[parse-numbers=false]{145.2(1.2)(1.4)(0.7)(4.7)} & \num[parse-numbers=false]{128.5(0.9)(1.9)(0.1)(0.5)} \\
      $\Re g$ & \num[parse-numbers=false]{5.937(12)(14)(20)(192)} & \num[parse-numbers=false]{5.765(51)(17)(0)(6)} \\
      $-\Im g$ & \num[parse-numbers=false]{0.7089(44)(51)(74)(230)} & \num[parse-numbers=false]{0.426(12)(10)(7)(9)} \\
    \end{tabular}
    \caption{Results of NLO and NNLO fits to the HadSpec data.
    The first error is the statistical one, the second the one due to the lattice spacing, the third arises from the error of the literature value of $F$, the final one is the truncation error.}\label{table:HadSpec_pole_positions}
  \end{ruledtabular}
\end{table}

In Ref.~\cite{Bolton:2015psa}, the NLO IAM was fit to the ensemble at $\mpi\approx\SI{236}{\MeV}$ only, yielding $\chi^2/\text{dof} = 1.26$ as well as $M_\rho = 755(2)(1)^{+20}_{-2}\,\si{\MeV}$ and $\Gamma_\rho=129(3)(1)^{+7}_{-1}\,\si{\MeV}$ at the physical point.
Comparison with our NLO fit (see Table~\ref{table:HadSpec_pole_positions}) shows that our $\chi^2/\text{dof}$ is significantly larger, but at the same time $\Gamma_\rho$ comes out much closer to its literature value $\Gamma_\rho = 146.4^{+2.0}_{-2.2}\,\si{\MeV}$~\cite{GarciaMartin:2011jx}.
The origin of these differences is twofold:
first, the NLO fit presented in this work is not restricted to one ensemble, but instead three ensembles are fit simultaneously.
Second, we express the ChPT partial waves in terms of $F$, while in Ref.~\cite{Bolton:2015psa} the amplitudes are expressed in terms of $F_\pi$, which contrary to $F$ depends on the pion mass, see App.~\ref{sec:appendix_pion_decay_chpt}. In consequence, the NLO expression for the $\pi\pi$ amplitude in Ref.~\cite{Bolton:2015psa} depends on both $l_2^r - 2l_1^r$ and $l_4^r$, but, as we argued here, the dependence on $l_4^r$ represents a spurious higher-order effect (even the two-loop IAM does not depend on $l_4^r$ once expressed in terms of $F$). Since both LECs are fit only to $\pi\pi$ data, this additional freedom improves the fit, but at the expense of adjusting $l_4^r$ not from the pion-mass dependence of $F_\pi$, but from higher-order effects in $\pi\pi$ scattering. The resulting value of $l_4^r = -28(6)(3)^{+1}_{-11} \times 10^{-3}$ indeed deviates considerably from its literature value, albeit in the same direction as we observed in the CLS fits, see App.~\ref{sec:appendix_cls_scale_setting}.

  \newpage
  \section{ChPT expressions}
\label{sec:ChPT}

\subsection{Strategy for partial-wave projection}

Starting point for the partial-wave projection is the $\pi\pi$ scattering amplitude in Ref.~\cite{Bijnens1996}. For the $1$-loop amplitudes the derivation is relatively straightforward: in this case, the integration of the loop functions naively gives rise to polylogarithms $\Li_2$, but the final result can be expressed solely in terms of the logarithm 
\beq
L(s)=\log\frac{1+\sigma(s)}{1-\sigma(s)},\qquad \sigma(s)=\sqrt{1-\frac{4\mpi^2}{s}}.
\eeq

For the two-loop amplitudes, we first observe that the complicated part of the partial-wave projection can be reduced to the three integrals
\beq
\label{proj_int}
\int_{-1}^1dz\, F^{(2)}(t)P_J(z),\qquad
\int_{-1}^1dz\, G^{(2)}(s,t)P_J(z),\qquad 
\int_{-1}^1dz\, G^{(2)}(u,t)P_J(z),
\eeq
where $F^{(2)}(t)$ and $G^{(2)}(s,t)$ are the loop functions as defined in Ref.~\cite{Bijnens1996}, with Mandelstam variables $s$, $t$, $u$ related to the scattering angle $z$ according to
\beq
t(z)=-2q^2(1-z),\qquad u(z)=-2q^2(1+z),\qquad q^2=\frac{s}{4}-\mpi^2.
\eeq
We then perform a crucial change of variables 
\beq
z\to \tilde t(z)=\frac{\sigma(t(z))+1}{\sigma(t(z))-1},\qquad
t(z)=-\frac{\mpi^2(1-\tilde t(z))^2}{\tilde t(z)},
\eeq
as motivated by the arguments of the $t$-channel loop functions,
which allows one to express the integrals in Eq.~\eqref{proj_int} as complicated functions of polylogarithms up to $\Li_3$. From this full set of basic integrals, a raw form of the entire amplitudes follows. However, the results obtained in this way are not yet useful, as extremely lengthy and involving delicate cancellations among complex-valued functions.

Next, we have systematically applied relations among polylogarithms with different arguments, e.g.,
\beq
\Li_2(z)+\Li_2(1-z)+\log z\log(1-z)=\frac{\pi^2}{6},
\eeq
to reduce the expressions to as few real-valued functions (in the physical region) as possible, which leads to $\Li_3(\sigma_\pm)$ and $\Li_2(\sigma_-)$ in Eq.~\eqref{two-loop}. Next, we converted all logarithms to real-valued functions, which in the end only leaves powers of $L(s)$. As cross checks of this procedure, we have verified numerically that our analytic results agree with the numerical form of the partial-wave projection. In addition, we extracted the imaginary parts of the two-loop amplitudes and checked analytically that they agree with the ones dictated by elastic unitarity, which for $S$- and $P$-waves amounts to a rederivation of the one-loop amplitudes via the imaginary part of the two-loop ones, while for $D$-waves the imaginary parts cancel in accord with the vanishing tree-level contributions.

\subsection{Partial waves at one-loop order}

We write the NLO partial-wave amplitudes as
\beq
\Re t_J^I(s)\big|_4= \sum_{i=0}^2 b_i^{IJ}(s) \left[L(s)\right]^i +  \sum_{i=1}^3 b_{l_i}^{IJ}(s) l_i^r.
\eeq
The coefficient functions are
\begin{align}
b_0^{00}(s)&=\frac{373 \mpi^4-190 \mpi^2 s+51 s^2-5 \left(31 \mpi^4-32 \mpi^2 s+10 s^2\right) \log \frac{\mpi^2}{\mu
   ^2}}{9216 \pi ^3 F^4},\notag\\
b_1^{00}(s)&=\frac{36 \mpi^6-303 \mpi^4 s+260\mpi^2 s^2-50 s^3}{9216 \pi ^3 F^4 s \sigma(s) },\qquad
b_2^{00}(s)=\frac{\mpi^4 \left(6 s-25 \mpi^2\right)}{1536 \pi ^3 F^4 \left(s-4 \mpi^2\right)},\notag\\
b_{l_1}^{00}(s)&=\frac{44 \mpi^4-40 \mpi^2 s+11 s^2}{48 \pi  F^4},\qquad
b_{l_2}^{00}(s)=\frac{28 \mpi^4-20 \mpi^2 s+7
  s^2}{48 \pi  F^4},\qquad
b_{l_3}^{00}(s)=\frac{5 \mpi^4}{16 \pi  F^4},\\
b_0^{20}(s)&=\frac{80 \mpi^4-122 \mpi^2 s+51 s^2-4 \left(31 \mpi^4-32 \mpi^2 s+10 s^2\right) \log \frac{\mpi^2}{\mu ^2}}{18432 \pi ^3
   F^4},\notag\\
b_1^{20}(s)&=\frac{72 \mpi^6-93 \mpi^4 s+52 \mpi^2
   s^2-10 s^3}{4608 \pi ^3 F^4 s \sigma(s) },\qquad b_2^{20}(s)=-\frac{\mpi^4 \left(\mpi^2+3 s\right)}{1536 \pi ^3 F^4 \left(s-4 \mpi^2\right)},\notag\\
b_{l_1}^{20}(s)&=\frac{4 \mpi^4-2 \mpi^2 s+s^2}{24 \pi  F^4},\qquad b_{l_2}^{20}(s)=\frac{8 \mpi^4-7 \mpi^2 s+2 s^2}{24 \pi 
   F^4},\qquad b_{l_3}^{20}(s)=\frac{ \mpi^4}{8 \pi  F^4},\\
b_0^{11}(s)&=
   -\frac{120\mpi^6-197 \mpi^4 s+61 \mpi^2 s^2-2 s^3}{27648 \pi ^3 F^4 \left(s-4 \mpi^2\right)},\notag\\
b_1^{11}(s)&=-\frac{64 \mpi^8-55
   \mpi^6 s+6 \mpi^4 s^2}{2304 \pi ^3 F^4 s \sigma(s)  \left(s-4 \mpi^2\right)},\qquad b_2^{11}(s)=-\frac{ \mpi^4 \left(6 \mpi^4+13 \mpi^2 s-3 s^2\right)}{1536 \pi ^3 F^4 \left(s-4 \mpi^2\right)^2},\notag\\
b_{l_1}^{11}(s)&=-2b_{l_2}^{11}(s)=\frac{s\left(4 \mpi^2-s\right)}{48 \pi  F^4},\qquad b_{l_3}^{11}(s)=0,\\
b_0^{02}(s)&=\frac{13376 \mpi^8-11946 \mpi^6 s+2481 \mpi^4 s^2-191 \mpi^2 s^3+6 s^4-10 \left(s-4 \mpi^2\right)^4 \log\frac{\mpi^2}{\mu^2}}{46080 \pi^3 F^4 \left(s-4 \mpi^2\right)^2},\notag\\
b_1^{02}(s)&=-\frac{516 \mpi^8-760 \mpi^6 s+180 \mpi^4 s^2-18 \mpi^2 s^3+s^4}{4608 \pi^3 F^4 \sigma(s)  \left(s-4 \mpi^2\right)^2},\qquad b_2^{02}(s)=-\frac{\mpi^4 \left(172 \mpi^6-98 \mpi^4 s+49 \mpi^2 s^2-6 s^3\right)}{1536 \pi ^3 F^4 \left(s-4 \mpi^2\right)^3},\notag\\
2b_{l_1}^{02}(s)&=b_{l_2}^{02}(s)=\frac{\left(s-4 \mpi^2\right)^2}{120 \pi  F^4},\qquad b_{l_3}^{02}(s)=0,\\
b_0^{22}(s)&=-\frac{22784 \mpi^8+2496 \mpi^6 s+10824 \mpi^4 s^2-554 \mpi^2 s^3+9 s^4+80 \left(s-4 \mpi^2\right)^4\log\frac{\mpi^2}{\mu^2}}{921600 \pi ^3 F^4 \left(s-4 \mpi^2\right)^2},\notag\\
b_1^{22}(s)&=\frac{480 \mpi^8+980 \mpi^6 s-117 \mpi^4 s^2+36 \mpi^2 s^3-2 s^4}{23040 \pi ^3 F^4 \sigma(s) \left(s-4 \mpi^2\right)^2},\qquad b_2^{22}(s)=\frac{\mpi^4 \left(32 \mpi^6-76 \mpi^4 s+11 \mpi^2 s^2-3 s^3\right)}{1536 \pi ^3 F^4 \left(s-4 \mpi^2\right)^3},\notag\\
b_{l_1}^{22}(s)&=2b_{l_2}^{22}(s)=\frac{\left(s-4 \mpi^2\right)^2}{240 \pi  F^4},\qquad b_{l_3}^{22}(s)=0.
\end{align}
In particular, there are no contributions involving $l_4^r$ when the amplitudes are written in terms of $F$ instead of $F_\pi$. The $P$-wave amplitude depends on a single combination of LECs, $2l_1^r-l_2^r$, which is scale independent.

\subsection{Partial waves at two-loop order}

We write the NNLO partial waves as
\beq
\label{two-loop}
\Re t_J^I(s)\big|_6=\sum_{i=0}^4c_i^{IJ}(s) \left[L(s)\right]^i+ \sum_{i=1}^3c_{l_i}^{IJ}(s) l_i^r+c_{l_3^2}^{IJ}(s)\big(l_3^r\big)^2
+d^{IJ}(s)\Big[\Li_3\big(\sigma_+(s)\big)+\Li_3\big(\sigma_-(s)\big)-L(s)\, \Li_2\big(\sigma_-(s)\big)\Big]
+P^{IJ}(s),
 \eeq
 with
\beq
\sigma_\pm(s)=\frac{2\sigma(s)}{\sigma(s)\pm 1},\qquad 
\Li_n(x)=\frac{(-1)^{n-1}}{(n-2)!}\int_0^1dt\frac{\log^{n-2}t\log(1-x t)}{t}=\sum_{k=1}^\infty\frac{x^k}{k^n},
\eeq
and where the two-loop LECs are contained in
\begin{align}
P^{I0}(s)&=\frac{1}{\pi F^6}\big(r_a^{I0} \mpi^6 + r_b^{I0}s\mpi^4+r_c^{I0}s^2\mpi^2+r_d^{I0}s^3\big),\notag\\
P^{11}(s)&=\frac{s-4\mpi^2}{\pi F^6}\big(r_a^{11} \mpi^4 + r_b^{11}s\mpi^2+r_c^{11}s^2\big),\notag\\
P^{I2}(s)&=\frac{(s-4\mpi^2)^2}{\pi F^6}\big(r_a^{I2} \mpi^2 + r_b^{I2}s\big),
\end{align}
with
\begin{align}
 r_a^{00}&=\frac{5r_1^r}{32}+\frac{r_2^r}{8}+\frac{r_3^r}{3}+\frac{5r_4^r}{6}+r_5^r+\frac{r_6^r}{3}+\frac{r_F^r}{4096\pi^4},\qquad
 r_b^{00}=\frac{r_2^r}{16}-\frac{r_3^r}{6}-\frac{2r_4^r}{3}-\frac{3r_5^r}{4}-\frac{r_6^r}{12}-\frac{r_F^r}{2048\pi^4},\notag\\
 r_c^{00}&=\frac{11r_3^r}{96}+\frac{17r_4^r}{96}+\frac{3r_5^r}{16}+\frac{5r_6^r}{48},\qquad r_d^{00}=\frac{5r_5^r}{64}-\frac{5r_6^r}{192},\notag\\
   r_a^{20}&=\frac{r_1^r}{16}+\frac{r_2^r}{8}+\frac{r_3^r}{3}+\frac{r_4^r}{3}+r_5^r+\frac{r_6^r}{3}-\frac{r_F^r}{2048\pi^4},\qquad
 r_b^{20}=-\frac{r_2^r}{32}-\frac{r_3^r}{6}-\frac{5r_4^r}{12}-\frac{3r_5^r}{4}-\frac{7r_6^r}{12}+\frac{r_F^r}{4096\pi^4},\notag\\
 r_c^{20}&=\frac{r_3^r}{48}+\frac{7r_4^r}{48}+\frac{3r_5^r}{16}+\frac{17r_6^r}{48},\qquad r_d^{20}=-\frac{r_5^r}{64}-\frac{11r_6^r}{192},\notag\\
 r_a^{11}&=\frac{r_2^r}{96}+\frac{r_3^r}{24}-\frac{r_4^r}{24}+\frac{3r_5^r}{20}-\frac{r_6^r}{60}-\frac{r_F^r}{12288\pi^4},\qquad
 r_b^{11}=-\frac{r_3^r}{96}+\frac{r_4^r}{32}-\frac{3r_5^r}{40}+\frac{r_6^r}{120},\qquad r_c^{11}=\frac{3}{320}(r_5^r+r_6^r),\notag\\
 r_a^{02}&=\frac{r_3^r}{480}+\frac{7r_4^r}{480}+\frac{r_5^r}{80}-\frac{r_6^r}{240},\qquad r_b^{02}=-\frac{r_5^r}{320}+\frac{17r_6^r}{960},\notag\\
 r_a^{22}&=\frac{r_3^r}{480}+\frac{r_4^r}{480}+\frac{r_5^r}{80}-\frac{r_6^r}{240},\qquad r_b^{22}=-\frac{r_5^r}{320}+\frac{5r_6^r}{960}.
\end{align}
The other coefficient functions are
\begin{align}
c_0^{00}(s)&=\frac{1}{127401984 \pi ^5 F^6 s}\Bigg[-15552 \pi ^2 \mpi^8+2 \left(90857+80856 \pi ^2\right) \mpi^6 s+4 \left(402103-76368 \pi ^2\right) \mpi^4 s^2\notag\\
&+4 \left(53808 \pi ^2-135623\right) \mpi^2 s^3+3 \left(22989-12880 \pi ^2\right) s^4+24 s  \log
\frac{\mpi^2}{\mu ^2} \bigg(3822 \mpi^6-54416 \mpi^4 s\notag\\
&+22120 \mpi^2 s^2-3845 s^3-60 \left(98 \mpi^6-319 \mpi^4 s+161 \mpi^2 s^2-30 s^3\right) \log
\frac{\mpi^2}{\mu ^2}\bigg)\Bigg],\notag\\
c_1^{00}(s)&=\frac{1}{5308416 \pi ^5 F^6 s \sigma(s) }\Bigg[144 \left(25 \pi ^2-373\right) \mpi^8+144 \left(911-66 \pi ^2\right) \mpi^6 s+6 \left(288 \pi ^2-16481\right) \mpi^4 s^2\notag\\
&+29810 \mpi^2 s^3-3845 s^4
+60 \left(372 \mpi^8-1107 \mpi^6
   s+1126 \mpi^4 s^2-442 \mpi^2 s^3+60 s^4\right)  \log
\frac{\mpi^2}{\mu ^2}\Bigg],\notag\\
c_2^{00}(s)&=\frac{1}{884736 \pi ^5 F^6 s \left(s-4 \mpi^2\right)}\Bigg[-432 \mpi^{10}+6 \left(295-18 \pi ^2\right) \mpi^8 s+\left(120 \pi ^2-12091\right) \mpi^6 s^2\notag\\
&+9774 \mpi^4 s^3-2810 \mpi^2 s^4+300 s^5+1140 \mpi^8 s \log\frac{\mpi^2}{\mu ^2}\Bigg],\notag\\
c_3^{00}(s)&=-\frac{\mpi^6 \left(53 \mpi^2+32 s\right)}{36864 \pi ^5 F^6 \left(s-4 \mpi^2\right)}-\frac{\mpi^4 \left(25 \mpi^4-36 \mpi^2 s+12 s^2\right)}{73728 \pi ^5 F^6 s \sigma(s) },\qquad
c_4^{00}(s)=\frac{\mpi^6 \left(10 s-9 \mpi^2\right)}{147456 \pi ^5 F^6 \left(s-4 \mpi^2\right)},\notag\\
c_{l_1}^{00}(s)&=-\frac{\mpi^6 \left(9 \mpi^2+4 s\right)}{384 \pi ^3 F^6 \left(s-4 \mpi^2\right)}\left[L(s)\right]^2-\frac{1056 \mpi^8-3228 \mpi^6 s+2898 \mpi^4 s^2-1000 \mpi^2 s^3+115
   s^4}{4608 \pi ^3 F^6 s \sigma(s) }L(s)\notag\\
   &+\frac{\left(726 \mpi^6-1804 \mpi^4 s+770 \mpi^2 s^2-115 s^3\right) \log \frac{\mpi^2}{\mu ^2}}{4608 \pi ^3 F^6}
   -\frac{324\mpi^6-3788 \mpi^4 s+1964 \mpi^2 s^2-447 s^3}{18432 \pi ^3 F^6},\notag\\
c_{l_2}^{00}(s)&=\frac{\mpi^6 \left(4 s-61 \mpi^2\right)}{768 \pi ^3 F^6 \left(s-4 \mpi^2\right)}\left[L(s)\right]^2-\frac{1344 \mpi^8-3252 \mpi^6 s+3002 \mpi^4 s^2-1210 \mpi^2 s^3+185
   s^4}{9216 \pi ^3 F^6 s \sigma(s) }L(s)\notag\\
   &-\frac{\left(16 \mpi^6+1656 \mpi^4 s-840 \mpi^2 s^2+185 s^3\right) \log \frac{\mpi^2}{\mu ^2}}{9216 \pi ^3 F^6}
   +\frac{4784\mpi^6+4612 \mpi^4 s-2756 \mpi^2 s^2+833 s^3}{36864 \pi ^3 F^6},\notag\\
 c_{l_3}^{00}(s)&=-\frac{5\mpi^8}{128 \pi ^3 F^6 \left(s-4 \mpi^2\right)}\left[L(s)\right]^2-\frac{5\left(4 \mpi^8-7 \mpi^6 s+2 \mpi^4 s^2\right)}{256 \pi ^3 F^6 s \sigma(s) }L(s)
 -\frac{5 \mpi^4
   \left(\mpi^2+s\right) \log\frac{\mpi^2}{\mu ^2}}{128 \pi ^3 F^6}+\frac{\mpi^4 \left(43 \mpi^2+4 s\right)}{512 \pi ^3 F^6},\notag\\
 c_{l_3^2}^{00}(s)&=-\frac{5 \mpi^6}{4 \pi  F^6},\qquad d^{00}(s)=\frac{\mpi^6 \left(53 \mpi^2+32 s\right)}{9216 \pi ^5 F^6 \left(s-4 \mpi^2\right)},\\
c_0^{20}(s)&=\frac{1}{254803968 \pi ^5 F^6 s}\Bigg[248832 \pi^2 \mpi^8+8 \left(112123-51840 \pi ^2\right) \mpi^6 s+4 \left(71544 \pi ^2-152179\right) \mpi^4 s^2\notag\\
&+4 \left(44861-26016 \pi ^2\right) \mpi^2 s^3+3 \left(4384 \pi ^2-22641\right) s^4
-24 s \log\frac{\mpi^2}{\mu ^2} \bigg(44280 \mpi^6-41588 \mpi^4 s\notag\\
&+16300 \mpi^2 s^2-3197 s^3-12 \left(1324 \mpi^6-1499 \mpi^4 s+640 \mpi^2 s^2-105 s^3\right) \log
   \frac{\mpi^2}{\mu ^2}\bigg)\Bigg],\notag\\
c_1^{20}(s)&=\frac{1}{10616832 \pi ^5 F^6 s \sigma(s) }\Bigg[576 \left(40-\pi ^2\right) \mpi^8-108 \left(691+32 \pi ^2\right) \mpi^6 s+30 \left(2459+72 \pi ^2\right) \mpi^4 s^2\notag\\
&-22694 \mpi^2 s^3+3197 s^4-24 \left(1488 \mpi^8-2844 \mpi^6 s+2479
   \mpi^4 s^2-850 \mpi^2 s^3+105 s^4\right)  \log
\frac{\mpi^2}{\mu ^2}\Bigg],\notag\\
c_2^{20}(s)&=\frac{1}{884736 \pi ^5 F^6 s \left(s-4 \mpi^2\right)}\Bigg[3456 \mpi^{10}+3 \left(12 \pi ^2-2969\right) \mpi^8 s+2 \left(4159-30 \pi ^2\right) \mpi^6 s^2\notag\\
&-4293 \mpi^4 s^3+1060 \mpi^2 s^4
-105 s^5+384 \mpi^8 s \log
   \frac{\mpi^2}{\mu ^2}\Bigg],\notag\\
c_3^{20}(s)&=\frac{\mpi^6 \left(16 s-41 \mpi^2\right)}{36864 \pi ^5 F^6 \left(s-4 \mpi^2\right)}+\frac{\mpi^4 \left(2 \mpi^4-9 \mpi^2 s+6 s^2\right)}{73728 \pi ^5 F^6 s \sigma(s) },\qquad
c_4^{20}(s)=\frac{\mpi^6 \left(3 \mpi^2-5 s\right)}{147456 \pi ^5 F^6 \left(s-4 \mpi^2\right)},\notag\\
c_{l_1}^{20}(s)&=\frac{\mpi^6 \left(4 s-15 \mpi^2\right)}{768 \pi ^3 F^6 \left(s-4 \mpi^2\right)}\left[L(s)\right]^2+\frac{768 \mpi^8-1140 \mpi^6 s+1122 \mpi^4 s^2-430 \mpi^2 s^3+61
   s^4}{9216 \pi ^3 F^6 s \sigma(s) }L(s)\notag\\
   &-\frac{\left(984 \mpi^6-844 \mpi^4 s+308 \mpi^2 s^2-61 s^3\right) \log \frac{\mpi^2}{\mu ^2}}{9216 \pi ^3 F^6}
   +\frac{3984\mpi^6-5084 \mpi^4 s+1796 \mpi^2 s^2-285 s^3}{36864 \pi ^3 F^6},\notag\\
c_{l_2}^{20}(s)&=-\frac{\mpi^6 \left(35 \mpi^2+4 s\right)}{1536 \pi ^3 F^6 \left(s-4 \mpi^2\right)}\left[L(s)\right]^2+\frac{3072 \mpi^8-5412 \mpi^6 s+4106 \mpi^4 s^2-1270 \mpi^2 s^3+149
   s^4}{18432 \pi ^3 F^6 s \sigma(s) }L(s)\notag\\
   &-\frac{\left(2528 \mpi^6-2424 \mpi^4 s+972 \mpi^2 s^2-149 s^3\right) \log \frac{\mpi^2}{\mu ^2}}{18432 \pi ^3 F^6}
   +\frac{8896\mpi^6-10084 \mpi^4 s+4100 \mpi^2 s^2-701 s^3}{73728 \pi ^3 F^6},\notag\\
 c_{l_3}^{20}(s)&=-\frac{\mpi^8}{64 \pi ^3 F^6 \left(s-4 \mpi^2\right)}\left[L(s)\right]^2+\frac{16 \mpi^8-16 \mpi^6 s+5 \mpi^4 s^2}{256 \pi ^3 F^6 s \sigma(s) }L(s)
 -\frac{5 \mpi^4 \left(8 \mpi^4-6
   \mpi^2 s+s^2\right)}{512 \pi ^3 F^6 \left(s-4 \mpi^2\right)}\notag\\
   &+\frac{\mpi^4 \left(5 s-16 \mpi^2\right) \log \frac{\mpi^2}{\mu ^2}}{256 \pi ^3 F^6},\notag\\
 c_{l_3^2}^{20}(s)&=-\frac{\mpi^6}{2 \pi  F^6},\qquad d^{20}(s)=\frac{\mpi^6 \left(41 \mpi^2-16 s\right)}{9216 \pi ^5 F^6 \left(s-4 \mpi^2\right)},\\
c_0^{11}(s)&=\frac{1}{6370099200 \pi ^5 F^6 s \left(s-4
   \mpi^2\right)}\Bigg[7372800 \pi ^2 \mpi^{10}+704 \left(178753-17280 \pi ^2\right) \mpi^8 s\notag\\
   &+128 \left(3765 \pi ^2-825499\right) \mpi^6 s^2+4 \left(7303133+246120 \pi ^2\right) \mpi^4 s^3-12
   \left(572531+28840 \pi ^2\right) \mpi^2 s^4\notag\\
   &+\left(654787+38880 \pi ^2\right) s^5-120 s  \log
\frac{\mpi^2}{\mu ^2} \bigg(714048 \mpi^8-518888 \mpi^6 s+229588 \mpi^4 s^2-65468 \mpi^2 s^3\notag\\
&+6633 s^4-300 \left(s-4
   \mpi^2\right)^2 \left(29 \mpi^4-10 \mpi^2 s+9 s^2\right)  \log
\frac{\mpi^2}{\mu ^2}\bigg)\Bigg],\notag\\
c_1^{11}(s)&=\frac{1}{53084160 \pi ^5 F^6 s \sigma(s) 
   \left(s-4 \mpi^2\right)}\Bigg[3840 \left(20+3 \pi ^2\right) \mpi^{10}+80 \left(1356 \pi ^2-8977\right) \mpi^8 s\notag\\
   &+20 \left(52063-1032 \pi ^2\right) \mpi^6 s^2+2 \left(360 \pi ^2-186289\right) \mpi^4 s^3+78734
   \mpi^2 s^4-6633 s^5\notag\\
   &+600 s \left(348 \mpi^8-710 \mpi^6 s+393 \mpi^4 s^2-100 \mpi^2 s^3+9 s^4\right)  \log
\frac{\mpi^2}{\mu ^2}\Bigg],\notag\\
c_2^{11}(s)&=\frac{1}{884736 \pi ^5 F^6 s \left(s-4 \mpi^2\right)^2}\Bigg[4096 \mpi^{12}-108 \left(137+4 \pi ^2\right) \mpi^{10} s+\left(10159-300 \pi ^2\right) \mpi^8 s^2\notag\\
&+60 \left(\pi ^2-133\right) \mpi^6 s^3+2995 \mpi^4 s^4-590 \mpi^2 s^5+45 s^6+120
   \mpi^8 s \left(29 \mpi^2-s\right)  \log
\frac{\mpi^2}{\mu ^2}\Bigg],\notag\\
c_3^{11}(s)&=\frac{\mpi^6 \left(95 \mpi^2-16 s\right)}{36864 \pi ^5 F^6 \left(s-4 \mpi^2\right)}-\frac{\mpi^4 \left(24 \mpi^6-314 \mpi^4 s+11 \mpi^2 s^2+6 s^3\right)}{221184 \pi ^5 F^6 s
   \sigma(s)  \left(s-4 \mpi^2\right)},\notag\\
c_4^{11}(s)&=-\frac{\mpi^6 \left(36 \mpi^4+25 \mpi^2 s-5 s^2\right)}{147456 \pi ^5 F^6 \left(s-4 \mpi^2\right)^2},\notag\\
c_{l_1}^{11}(s)&=-\frac{\mpi^6 \left(116 \mpi^4-11 \mpi^2 s+4 s^2\right)}{768 \pi ^3 F^6 \left(s-4 \mpi^2\right)^2}\left[L(s)\right]^2
-\frac{1840 \mpi^8-8780 \mpi^6 s+5498 \mpi^4 s^2-1594
   \mpi^2 s^3+153 s^4}{46080 \pi ^3 F^6 \sigma(s)  \left(s-4 \mpi^2\right)}L(s)\notag\\
   &-\frac{\left(s-4 \mpi^2\right) \left(628 \mpi^4-64 \mpi^2 s+153 s^2\right) \log\frac{\mpi^2}{\mu ^2}}{46080 \pi ^3 F^6}\notag\\
   &+\frac{1213056 \mpi^8-1003536 \mpi^6 s+426916 \mpi^4 s^2-112736 \mpi^2 s^3+10531 s^4}{2764800 \pi ^3 F^6 \left(s-4
   \mpi^2\right)},\notag\\
c_{l_2}^{11}(s)&=\frac{ \mpi^6 \left(8 \mpi^2+s\right) \left(4 s-33 \mpi^2\right)}{1536 \pi ^3 F^6 \left(s-4 \mpi^2\right)^2}\left[L(s)\right]^2
-\frac{20960 \mpi^8-31220 \mpi^6 s+14602 \mpi^4
   s^2-3406 \mpi^2 s^3+297 s^4}{92160 \pi ^3 F^6 \sigma(s)  \left(s-4 \mpi^2\right)}L(s)\notag\\
   &-\frac{\left(s-4 \mpi^2\right) \left(1272 \mpi^4-436 \mpi^2 s+297 s^2\right) \log
   \frac{\mpi^2}{\mu ^2}}{92160 \pi ^3 F^6}\notag\\
   &+\frac{2694144 \mpi^8-2309664 \mpi^6 s+984284 \mpi^4 s^2-240964 \mpi^2 s^3+21419 s^4}{5529600 \pi ^3 F^6 \left(s-4
   \mpi^2\right)},\notag\\
 c_{l_3}^{11}(s)&=-\frac{5\mpi^{10}}{64 \pi ^3 F^6 \left(s-4 \mpi^2\right)^2}\left[L(s)\right]^2-\frac{5\mpi^4 \left(12 \mpi^4-10 \mpi^2 s+s^2\right)}{768 \pi ^3 F^6 \sigma(s)  \left(s-4
   \mpi^2\right)}L(s)
   +\frac{\mpi^4 \left(496 \mpi^4-218 \mpi^2 s+s^2\right)}{4608 \pi ^3 F^6 \left(s-4 \mpi^2\right)}\notag\\
   &-\frac{5 \mpi^4 \left(s-4 \mpi^2\right) \log
   \frac{\mpi^2}{\mu ^2}}{768 \pi ^3 F^6},\notag\\
 c_{l_3^2}^{11}(s)&=0,\qquad d^{11}(s)=\frac{\mpi^6 \left(16 s-95 \mpi^2\right)}{9216 \pi ^5 F^6 \left(s-4 \mpi^2\right)},\\
c_0^{02}(s)&=\frac{1}{3185049600 \pi ^5 F^6 \left(s-4
   \mpi^2\right)^2}\Bigg[2048 \left(45323+3240 \pi ^2\right) \mpi^{10}-64 \left(430069+171975 \pi ^2\right) \mpi^8 s\notag\\
   &+96 \left(8900 \pi ^2-545299\right) \mpi^6 s^2+176 \left(66889+2700 \pi ^2\right) \mpi^4 s^3-4
   \left(218861+13200 \pi ^2\right) \mpi^2 s^4\notag\\
   &+3 \left(11861+1040 \pi ^2\right) s^5-120 \log \frac{\mpi^2}{\mu ^2} \bigg(496896 \mpi^{10}-264104 \mpi^8 s-38356 \mpi^6 s^2+20916 \mpi^4 s^3\notag\\
   &-3046\mpi^2 s^4+157 s^5-300 \mpi^2 \left(s-4 \mpi^2\right)^4 \log \frac{\mpi^2}{\mu ^2}\bigg)\Bigg],\notag\\
c_1^{02}(s)&=\frac{1}{26542080 \pi ^5 F^6 \sigma(s)  \left(s-4 \mpi^2\right)^2}\Bigg[-1440 \left(623+17 \pi ^2\right) \mpi^{10}+240 \left(4159+441 \pi ^2\right) \mpi^8 s\notag\\
&+4 \left(19081-5760 \pi ^2\right) \mpi^6 s^2-40734 \mpi^4 s^3+3360 \mpi^2 s^4-157 s^5\notag\\
&+600 \mpi^2
   \left(444 \mpi^8-580 \mpi^6 s+126 \mpi^4 s^2-18 \mpi^2 s^3+s^4\right) \log \frac{\mpi^2}{\mu ^2}\Bigg],\notag\\
c_2^{02}(s)&=\frac{\mpi^2}{442368 \pi ^5 F^6 \left(s-4 \mpi^2\right)^3}\Bigg[24 \left(27 \pi ^2-623\right) \mpi^{10}+8 \left(1487-105 \pi ^2\right) \mpi^8 s-\left(7541+102 \pi ^2\right) \mpi^6 s^2\notag\\
&+12 \left(71+5 \pi ^2\right) \mpi^4 s^3-46
   \mpi^2 s^4+5 s^5+30 \mpi^6 \left(148 \mpi^4-50 \mpi^2 s+19 s^2\right) \log \frac{\mpi^2}{\mu ^2}\Bigg],\notag\\
c_3^{02}(s)&=-\frac{\mpi^6 \left(53 \mpi^2+32 s\right)}{36864 \pi ^5 F^6 \left(s-4 \mpi^2\right)}-\frac{\mpi^6 \left(34 \mpi^4-147 \mpi^2 s+32 s^2\right)}{36864 \pi ^5 F^6 \sigma(s)  \left(s-4
   \mpi^2\right)^2},\notag\\
c_4^{02}(s)&=\frac{\mpi^6 \left(108 \mpi^6-140 \mpi^4 s-17 \mpi^2 s^2+10 s^3\right)}{147456 \pi ^5 F^6 \left(s-4 \mpi^2\right)^3},\notag\\
c_{l_1}^{02}(s)&=-\frac{\mpi^6 \left(72 \mpi^6-26 \mpi^4 s+13 \mpi^2 s^2+4 s^3\right)}{384 \pi ^3 F^6 \left(s-4 \mpi^2\right)^3}\left[L(s)\right]^2-\frac{\left(6 \mpi^2-13 s\right) \left(s-4 \mpi^2\right)^2 \log
   \frac{\mpi^2}{\mu ^2}}{23040 \pi ^3 F^6}\notag\\
&-\frac{4320 \mpi^{10}-10440 \mpi^8 s+4696
   \mpi^6 s^2-1746 \mpi^4 s^3+240 \mpi^2 s^4-13 s^5}{23040 \pi ^3 F^6 \sigma(s)  \left(s-4 \mpi^2\right)^2}L(s)\notag\\
   &+\frac{282624 \mpi^{10}-314976 \mpi^8 s+116056 \mpi^6 s^2-33996 \mpi^4 s^3+3666 \mpi^2 s^4-157 s^5}{460800 \pi ^3 F^6
   \left(s-4 \mpi^2\right)^2},\notag\\
c_{l_2}^{02}(s)&=-\frac{ \mpi^6 \left(388 \mpi^6-144 \mpi^4 s+57 \mpi^2 s^2-4 s^3\right)}{768 \pi ^3 F^6 \left(s-4 \mpi^2\right)^3}\left[L(s)\right]^2-\frac{\left(s-4 \mpi^2\right)^2 \left(44 \mpi^2+13 s\right) \log
   \frac{\mpi^2}{\mu ^2}}{46080 \pi ^3 F^6}\notag\\
&-\frac{23280 \mpi^{10}-23960 \mpi^8 s+1604
   \mpi^6 s^2+846 \mpi^4 s^3-190 \mpi^2 s^4+13 s^5}{46080 \pi ^3 F^6 \sigma(s)  \left(s-4 \mpi^2\right)^2}L(s)\notag\\
   &+\frac{1150976 \mpi^{10}-680624 \mpi^8 s+10544 \mpi^6 s^2+23896 \mpi^4 s^3-3316 \mpi^2 s^4+157 s^5}{921600 \pi ^3 F^6
   \left(s-4 \mpi^2\right)^2},\notag\\
 c_{l_3}^{02}(s)&=-\frac{5  \mpi^8 \left(4 \mpi^4-2 \mpi^2 s+s^2\right)}{128 \pi ^3 F^6 \left(s-4 \mpi^2\right)^3}\left[L(s)\right]^2+\frac{5 \mpi^8 \left(s-\mpi^2\right)}{32 \pi ^3 F^6 \sigma(s)  \left(s-4
   \mpi^2\right)^2}L(s)+\frac{5 \mpi^6 \left(64 \mpi^4-38 \mpi^2 s+s^2\right)}{768 \pi ^3 F^6 \left(s-4 \mpi^2\right)^2},\notag\\
 c_{l_3^2}^{02}(s)&=0,\qquad d^{02}(s)=\frac{\mpi^6 \left(53 \mpi^2+32 s\right)}{9216 \pi ^5 F^6 \left(s-4 \mpi^2\right)},\\
c_0^{22}(s)&=\frac{1}{31850496000 \pi
   ^5 F^6 \left(s-4 \mpi^2\right)^2}\Bigg[2048 \left(93743-7200 \pi ^2\right) \mpi^{10}+64 \left(1483800 \pi ^2-2420311\right) \mpi^8 s\notag\\
   &+384 \left(322661-28675 \pi ^2\right) \mpi^6 s^2-8 \left(5203013+314100 \pi ^2\right) \mpi^4
   s^3+8 \left(479273+50100 \pi ^2\right) \mpi^2 s^4\notag\\
   &-15 \left(13553+1760 \pi ^2\right) s^5+120 \log \frac{\mpi^2}{\mu ^2} \bigg(1304064 \mpi^{10}+2865896 \mpi^8 s-1380176 \mpi^6 s^2\notag\\
   &+487146 \mpi^4
   s^3-73616 \mpi^2 s^4+4205 s^5+60 \left(14 \mpi^2-45 s\right) \left(s-4 \mpi^2\right)^4 \log \frac{\mpi^2}{\mu ^2}\bigg)\Bigg],\notag\\
c_1^{22}(s)&=\frac{1}{265420800 \pi ^5 F^6 \sigma(s) 
   \left(s-4 \mpi^2\right)^2}\Bigg[-1440 \left(2697+560 \pi ^2\right) \mpi^{10}+23760 \left(187-20 \pi ^2\right) \mpi^8 s\notag\\
   &+4 \left(28800 \pi ^2-623489\right) \mpi^6 s^2+711018 \mpi^4 s^3-82026 \mpi^2 s^4+4205 s^5\notag\\
   &-120
   \left(2100 \mpi^{10}+43850 \mpi^8 s-20664 \mpi^6 s^2+5922 \mpi^4 s^3-824 \mpi^2 s^4+45 s^5\right) \log \frac{\mpi^2}{\mu ^2}\Bigg],\notag\\
c_2^{22}(s)&=\frac{1}{4423680 \pi ^5 F^6 \left(s-4 \mpi^2\right)^3}\Bigg[72 \left(40 \pi ^2-899\right) \mpi^{12}+40 \left(613+150 \pi ^2\right) \mpi^{10} s\notag\\
&+5 \left(84 \pi ^2-8203\right) \mpi^8 s^2+60 \left(462-5 \pi ^2\right) \mpi^6 s^3-7615 \mpi^4
   s^4+914 \mpi^2 s^5-45 s^6\notag\\
   &-120 \mpi^8 \left(35 \mpi^4-199 \mpi^2 s-16 s^2\right) \log \frac{\mpi^2}{\mu ^2}\Bigg],\notag\\
c_3^{22}(s)&=\frac{\mpi^6 \left(16 s-41 \mpi^2\right)}{36864 \pi ^5 F^6 \left(s-4 \mpi^2\right)}-\frac{\mpi^6 \left(56 \mpi^4+33 \mpi^2 s-8 s^2\right)}{18432 \pi ^5 F^6 \sigma(s)  \left(s-4
   \mpi^2\right)^2},\notag\\
c_4^{22}(s)&=\frac{\mpi^6 \left(48 \mpi^6+100 \mpi^4 s+7 \mpi^2 s^2-5 s^3\right)}{147456 \pi ^5 F^6 \left(s-4 \mpi^2\right)^3},\notag\\
c_{l_1}^{22}(s)&=\frac{ \mpi^6 \left(84 \mpi^6-176 \mpi^4 s-11 \mpi^2 s^2+4 s^3\right)}{768 \pi ^3 F^6 \left(s-4 \mpi^2\right)^3}\left[L(s)\right]^2+\frac{\left(s-4 \mpi^2\right)^2 \left(12 \mpi^2+41 s\right) \log
   \frac{\mpi^2}{\mu ^2}}{46080 \pi ^3 F^6}\notag\\
&+\frac{5040 \mpi^{10}+32520 \mpi^8 s-17228
   \mpi^6 s^2+4950 \mpi^4 s^3-726 \mpi^2 s^4+41 s^5}{46080 \pi ^3 F^6 \sigma(s)  \left(s-4 \mpi^2\right)^2}L(s)\notag\\
   &-\frac{73728 \mpi^{10}+567696 \mpi^8 s-300176 \mpi^6 s^2+88536 \mpi^4 s^3-12756 \mpi^2 s^4+689 s^5}{921600 \pi ^3 F^6
   \left(s-4 \mpi^2\right)^2},\notag\\
c_{l_2}^{22}(s)&=\frac{ \mpi^6 \left(76 \mpi^6-348 \mpi^4 s-39 \mpi^2 s^2-4 s^3\right)}{1536 \pi ^3 F^6 \left(s-4 \mpi^2\right)^3}\left[L(s)\right]^2-\frac{\left(40 \mpi^2-49 s\right) \left(s-4 \mpi^2\right)^2 \log
   \frac{\mpi^2}{\mu ^2}}{92160 \pi ^3 F^6}\notag\\
&+\frac{4560 \mpi^{10}+64360 \mpi^8 s-24100
   \mpi^6 s^2+6894 \mpi^4 s^3-922 \mpi^2 s^4+49 s^5}{92160 \pi ^3 F^6 \sigma(s)  \left(s-4 \mpi^2\right)^2}L(s)\notag\\
   &+\frac{163840 \mpi^{10}-1148656 \mpi^8 s+403936 \mpi^6 s^2-124816 \mpi^4 s^3+16336 \mpi^2 s^4-841 s^5}{1843200 \pi ^3 F^6
   \left(s-4 \mpi^2\right)^2},\notag\\
 c_{l_3}^{22}(s)&=\frac{ \mpi^8 \left(5 \mpi^4-7 \mpi^2 s-s^2\right)}{64 \pi ^3 F^6 \left(s-4 \mpi^2\right)^3}\left[L(s)\right]^2+\frac{\mpi^8 \left(10 \mpi^2+17 s\right)}{128 \pi ^3 F^6 \sigma(s)  \left(s-4
   \mpi^2\right)^2}L(s)
   -\frac{\mpi^4 \left(256 \mpi^6+196 \mpi^4 s+64 \mpi^2 s^2-3 s^3\right)}{3072 \pi ^3 F^6 \left(s-4 \mpi^2\right)^2},\notag\\
 c_{l_3^2}^{22}(s)&=0,\qquad d^{22}(s)=\frac{\mpi^6 \left(41 \mpi^2-16 s\right)}{9216 \pi ^5 F^6 \left(s-4 \mpi^2\right)}.
\end{align}
Even at two-loop level there are no contributions involving $l_4^r$, while the higher powers in $l_3^r$ arise from the quark-mass renormalization of the pion mass.

\subsection{Pion decay constant}\label{sec:appendix_pion_decay_chpt}
The expression for $F_\pi$ to NNLO in ChPT is given in Ref.~\cite{Bijnens:1997vq}, see also Ref.~\cite{Bijnens:1998fm}.
Rewriting it in terms of $F$ and $M_\pi$ results in:
\begin{equation} \label{eq:pion_decay_constant_chpt}
    F_\pi
    = F\left[1 + F_4 \frac{\mpi^2}{16\pi^2 F^2} + F_6 \bigg(\frac{\mpi^2}{16\pi^2 F^2}\bigg)^2\right],
\end{equation}
with
\begin{align}
F_4 &= 16\pi^2 l_4^r - \log\frac{\mpi^2}{\mu^2}, \notag\\
F_6 &= \left(16\pi^2\right)^2r_F^r - 16\pi^2\left(l_2^r + \frac{1}{2}l_1^r + 32\pi^2 l_3^rl_4^r\right) - \frac{13}{192} 
        + \bigg( 16\pi^2\left(7l_1^r + 4l_2^r - l_4^r\right) + \frac{29}{12}\bigg)\log\frac{\mpi^2}{\mu^2} -\frac{3}{4}\log^2\frac{\mpi^2}{\mu^2}.
\end{align}

  \newpage  
\end{widetext}

\bibliography{literature.bib} 

\begin{thebibliography}{85}
\expandafter\ifx\csname natexlab\endcsname\relax\def\natexlab#1{#1}\fi
\expandafter\ifx\csname bibnamefont\endcsname\relax
  \def\bibnamefont#1{#1}\fi
\expandafter\ifx\csname bibfnamefont\endcsname\relax
  \def\bibfnamefont#1{#1}\fi
\expandafter\ifx\csname citenamefont\endcsname\relax
  \def\citenamefont#1{#1}\fi
\expandafter\ifx\csname url\endcsname\relax
  \def\url#1{\texttt{#1}}\fi
\expandafter\ifx\csname urlprefix\endcsname\relax\def\urlprefix{URL }\fi
\providecommand{\bibinfo}[2]{#2}
\providecommand{\eprint}[2][]{\url{#2}}

\bibitem[{\citenamefont{Brice{\~{n}}o et~al.}(2018)\citenamefont{Brice{\~{n}}o,
  Dudek, and Young}}]{Briceo2018}
\bibinfo{author}{\bibfnamefont{R.~A.} \bibnamefont{Brice{\~{n}}o}},
  \bibinfo{author}{\bibfnamefont{J.~J.} \bibnamefont{Dudek}}, \bibnamefont{and}
  \bibinfo{author}{\bibfnamefont{R.~D.} \bibnamefont{Young}},
  \bibinfo{journal}{Rev. Mod. Phys.} \textbf{\bibinfo{volume}{90}},
  \bibinfo{pages}{025001} (\bibinfo{year}{2018}), \eprint{1706.06223}.

\bibitem[{\citenamefont{Weinberg}(1979)}]{Weinberg:1978kz}
\bibinfo{author}{\bibfnamefont{S.}~\bibnamefont{Weinberg}},
  \bibinfo{journal}{Physica A} \textbf{\bibinfo{volume}{96}},
  \bibinfo{pages}{327} (\bibinfo{year}{1979}).

\bibitem[{\citenamefont{Gasser and Leutwyler}(1984)}]{Gasser1984}
\bibinfo{author}{\bibfnamefont{J.}~\bibnamefont{Gasser}} \bibnamefont{and}
  \bibinfo{author}{\bibfnamefont{H.}~\bibnamefont{Leutwyler}},
  \bibinfo{journal}{Annals Phys.} \textbf{\bibinfo{volume}{158}},
  \bibinfo{pages}{142} (\bibinfo{year}{1984}).

\bibitem[{\citenamefont{Gasser and Leutwyler}(1985)}]{Gasser:1984gg}
\bibinfo{author}{\bibfnamefont{J.}~\bibnamefont{Gasser}} \bibnamefont{and}
  \bibinfo{author}{\bibfnamefont{H.}~\bibnamefont{Leutwyler}},
  \bibinfo{journal}{Nucl. Phys. B} \textbf{\bibinfo{volume}{250}},
  \bibinfo{pages}{465} (\bibinfo{year}{1985}).

\bibitem[{\citenamefont{Aoki et~al.}(2007)}]{Aoki:2007rd}
\bibinfo{author}{\bibfnamefont{S.}~\bibnamefont{Aoki}} \bibnamefont{et~al.}
  (\bibinfo{collaboration}{CP-PACS}), \bibinfo{journal}{Phys. Rev. D}
  \textbf{\bibinfo{volume}{76}}, \bibinfo{pages}{094506}
  (\bibinfo{year}{2007}), \eprint{0708.3705}.

\bibitem[{\citenamefont{G{\"o}ckeler et~al.}(2008)\citenamefont{G{\"o}ckeler,
  Horsley, Nakamura, Pleiter, Rakow, Schierholz, and
  Zanotti}}]{Gockeler:2008kc}
\bibinfo{author}{\bibfnamefont{M.}~\bibnamefont{G{\"o}ckeler}},
  \bibinfo{author}{\bibfnamefont{R.}~\bibnamefont{Horsley}},
  \bibinfo{author}{\bibfnamefont{Y.}~\bibnamefont{Nakamura}},
  \bibinfo{author}{\bibfnamefont{D.}~\bibnamefont{Pleiter}},
  \bibinfo{author}{\bibfnamefont{P.~E.~L.} \bibnamefont{Rakow}},
  \bibinfo{author}{\bibfnamefont{G.}~\bibnamefont{Schierholz}},
  \bibnamefont{and} \bibinfo{author}{\bibfnamefont{J.}~\bibnamefont{Zanotti}}
  (\bibinfo{collaboration}{QCDSF}), \bibinfo{journal}{PoS}
  \textbf{\bibinfo{volume}{LATTICE2008}}, \bibinfo{pages}{136}
  (\bibinfo{year}{2008}), \eprint{0810.5337}.

\bibitem[{\citenamefont{Feng et~al.}(2011)\citenamefont{Feng, Jansen, and
  Renner}}]{Feng:2010es}
\bibinfo{author}{\bibfnamefont{X.}~\bibnamefont{Feng}},
  \bibinfo{author}{\bibfnamefont{K.}~\bibnamefont{Jansen}}, \bibnamefont{and}
  \bibinfo{author}{\bibfnamefont{D.~B.} \bibnamefont{Renner}},
  \bibinfo{journal}{Phys. Rev. D} \textbf{\bibinfo{volume}{83}},
  \bibinfo{pages}{094505} (\bibinfo{year}{2011}), \eprint{1011.5288}.

\bibitem[{\citenamefont{Lang et~al.}(2011)\citenamefont{Lang, Mohler,
  Prelovsek, and Vidmar}}]{Lang:2011mn}
\bibinfo{author}{\bibfnamefont{C.~B.} \bibnamefont{Lang}},
  \bibinfo{author}{\bibfnamefont{D.}~\bibnamefont{Mohler}},
  \bibinfo{author}{\bibfnamefont{S.}~\bibnamefont{Prelovsek}},
  \bibnamefont{and} \bibinfo{author}{\bibfnamefont{M.}~\bibnamefont{Vidmar}},
  \bibinfo{journal}{Phys. Rev. D} \textbf{\bibinfo{volume}{84}},
  \bibinfo{pages}{054503} (\bibinfo{year}{2011}), \bibinfo{note}{[Erratum:
  Phys. Rev. D {\bf 89}, 059903 (2014)]}, \eprint{1105.5636}.

\bibitem[{\citenamefont{Pelissier and Alexandru}(2013)}]{Pelissier:2012pi}
\bibinfo{author}{\bibfnamefont{C.}~\bibnamefont{Pelissier}} \bibnamefont{and}
  \bibinfo{author}{\bibfnamefont{A.}~\bibnamefont{Alexandru}},
  \bibinfo{journal}{Phys. Rev. D} \textbf{\bibinfo{volume}{87}},
  \bibinfo{pages}{014503} (\bibinfo{year}{2013}), \eprint{1211.0092}.

\bibitem[{\citenamefont{Metivet}(2015)}]{Metivet:2014bga}
\bibinfo{author}{\bibfnamefont{T.}~\bibnamefont{Metivet}}
  (\bibinfo{collaboration}{Budapest-Marseille-Wuppertal}),
  \bibinfo{journal}{PoS} \textbf{\bibinfo{volume}{LATTICE2014}},
  \bibinfo{pages}{079} (\bibinfo{year}{2015}), \eprint{1410.8447}.

\bibitem[{\citenamefont{Feng et~al.}(2015)\citenamefont{Feng, Aoki, Hashimoto,
  and Kaneko}}]{Feng:2014gba}
\bibinfo{author}{\bibfnamefont{X.}~\bibnamefont{Feng}},
  \bibinfo{author}{\bibfnamefont{S.}~\bibnamefont{Aoki}},
  \bibinfo{author}{\bibfnamefont{S.}~\bibnamefont{Hashimoto}},
  \bibnamefont{and} \bibinfo{author}{\bibfnamefont{T.}~\bibnamefont{Kaneko}},
  \bibinfo{journal}{Phys. Rev. D} \textbf{\bibinfo{volume}{91}},
  \bibinfo{pages}{054504} (\bibinfo{year}{2015}), \eprint{1412.6319}.

\bibitem[{\citenamefont{Wilson et~al.}(2015)\citenamefont{Wilson, Brice{\~n}o,
  Dudek, Edwards, and Thomas}}]{Wilson2015}
\bibinfo{author}{\bibfnamefont{D.~J.} \bibnamefont{Wilson}},
  \bibinfo{author}{\bibfnamefont{R.~A.} \bibnamefont{Brice{\~n}o}},
  \bibinfo{author}{\bibfnamefont{J.~J.} \bibnamefont{Dudek}},
  \bibinfo{author}{\bibfnamefont{R.~G.} \bibnamefont{Edwards}},
  \bibnamefont{and} \bibinfo{author}{\bibfnamefont{C.~E.}
  \bibnamefont{Thomas}}, \bibinfo{journal}{Phys. Rev. D}
  \textbf{\bibinfo{volume}{92}}, \bibinfo{pages}{094502}
  (\bibinfo{year}{2015}), \eprint{1507.02599}.

\bibitem[{\citenamefont{Bali et~al.}(2016)\citenamefont{Bali, Collins, Cox,
  Donald, G{\"o}ckeler, Lang, and Sch{\"a}fer}}]{Bali:2015gji}
\bibinfo{author}{\bibfnamefont{G.~S.} \bibnamefont{Bali}},
  \bibinfo{author}{\bibfnamefont{S.}~\bibnamefont{Collins}},
  \bibinfo{author}{\bibfnamefont{A.}~\bibnamefont{Cox}},
  \bibinfo{author}{\bibfnamefont{G.}~\bibnamefont{Donald}},
  \bibinfo{author}{\bibfnamefont{M.}~\bibnamefont{G{\"o}ckeler}},
  \bibinfo{author}{\bibfnamefont{C.~B.} \bibnamefont{Lang}}, \bibnamefont{and}
  \bibinfo{author}{\bibfnamefont{A.}~\bibnamefont{Sch{\"a}fer}}
  (\bibinfo{collaboration}{RQCD}), \bibinfo{journal}{Phys. Rev. D}
  \textbf{\bibinfo{volume}{93}}, \bibinfo{pages}{054509}
  (\bibinfo{year}{2016}), \eprint{1512.08678}.

\bibitem[{\citenamefont{Guo et~al.}(2016)\citenamefont{Guo, Alexandru, Molina,
  and D{\"o}ring}}]{Guo:2016zos}
\bibinfo{author}{\bibfnamefont{D.}~\bibnamefont{Guo}},
  \bibinfo{author}{\bibfnamefont{A.}~\bibnamefont{Alexandru}},
  \bibinfo{author}{\bibfnamefont{R.}~\bibnamefont{Molina}}, \bibnamefont{and}
  \bibinfo{author}{\bibfnamefont{M.}~\bibnamefont{D{\"o}ring}},
  \bibinfo{journal}{Phys. Rev. D} \textbf{\bibinfo{volume}{94}},
  \bibinfo{pages}{034501} (\bibinfo{year}{2016}), \eprint{1605.03993}.

\bibitem[{\citenamefont{Fu and Wang}(2016)}]{Fu:2016itp}
\bibinfo{author}{\bibfnamefont{Z.}~\bibnamefont{Fu}} \bibnamefont{and}
  \bibinfo{author}{\bibfnamefont{L.}~\bibnamefont{Wang}},
  \bibinfo{journal}{Phys. Rev. D} \textbf{\bibinfo{volume}{94}},
  \bibinfo{pages}{034505} (\bibinfo{year}{2016}), \eprint{1608.07478}.

\bibitem[{\citenamefont{Alexandrou et~al.}(2017)\citenamefont{Alexandrou,
  Leskovec, Meinel, Negele, Paul, Petschlies, Pochinsky, Rendon, and
  Syritsyn}}]{Alexandrou:2017mpi}
\bibinfo{author}{\bibfnamefont{C.}~\bibnamefont{Alexandrou}},
  \bibinfo{author}{\bibfnamefont{L.}~\bibnamefont{Leskovec}},
  \bibinfo{author}{\bibfnamefont{S.}~\bibnamefont{Meinel}},
  \bibinfo{author}{\bibfnamefont{J.}~\bibnamefont{Negele}},
  \bibinfo{author}{\bibfnamefont{S.}~\bibnamefont{Paul}},
  \bibinfo{author}{\bibfnamefont{M.}~\bibnamefont{Petschlies}},
  \bibinfo{author}{\bibfnamefont{A.}~\bibnamefont{Pochinsky}},
  \bibinfo{author}{\bibfnamefont{G.}~\bibnamefont{Rendon}}, \bibnamefont{and}
  \bibinfo{author}{\bibfnamefont{S.}~\bibnamefont{Syritsyn}},
  \bibinfo{journal}{Phys. Rev. D} \textbf{\bibinfo{volume}{96}},
  \bibinfo{pages}{034525} (\bibinfo{year}{2017}), \eprint{1704.05439}.

\bibitem[{\citenamefont{Andersen et~al.}(2019)\citenamefont{Andersen, Bulava,
  H{\"o}rz, and Morningstar}}]{Andersen2019}
\bibinfo{author}{\bibfnamefont{C.}~\bibnamefont{Andersen}},
  \bibinfo{author}{\bibfnamefont{J.}~\bibnamefont{Bulava}},
  \bibinfo{author}{\bibfnamefont{B.}~\bibnamefont{H{\"o}rz}}, \bibnamefont{and}
  \bibinfo{author}{\bibfnamefont{C.}~\bibnamefont{Morningstar}},
  \bibinfo{journal}{Nucl. Phys. B} \textbf{\bibinfo{volume}{939}},
  \bibinfo{pages}{145} (\bibinfo{year}{2019}), \eprint{1808.05007}.

\bibitem[{\citenamefont{Werner et~al.}(2020)}]{Werner:2019hxc}
\bibinfo{author}{\bibfnamefont{M.}~\bibnamefont{Werner}} \bibnamefont{et~al.}
  (\bibinfo{collaboration}{Extended Twisted Mass}), \bibinfo{journal}{Eur.
  Phys. J. A} \textbf{\bibinfo{volume}{56}}, \bibinfo{pages}{61}
  (\bibinfo{year}{2020}), \eprint{1907.01237}.

\bibitem[{\citenamefont{Erben et~al.}(2020)\citenamefont{Erben, Green, Mohler,
  and Wittig}}]{Erben:2019nmx}
\bibinfo{author}{\bibfnamefont{F.}~\bibnamefont{Erben}},
  \bibinfo{author}{\bibfnamefont{J.~R.} \bibnamefont{Green}},
  \bibinfo{author}{\bibfnamefont{D.}~\bibnamefont{Mohler}}, \bibnamefont{and}
  \bibinfo{author}{\bibfnamefont{H.}~\bibnamefont{Wittig}},
  \bibinfo{journal}{Phys. Rev. D} \textbf{\bibinfo{volume}{101}},
  \bibinfo{pages}{054504} (\bibinfo{year}{2020}), \eprint{1910.01083}.

\bibitem[{\citenamefont{Fischer et~al.}(2020)\citenamefont{Fischer, Kostrzewa,
  Mai, Petschlies, Pittler, Ueding, Urbach, and Werner}}]{Fischer:2020fvl}
\bibinfo{author}{\bibfnamefont{M.}~\bibnamefont{Fischer}},
  \bibinfo{author}{\bibfnamefont{B.}~\bibnamefont{Kostrzewa}},
  \bibinfo{author}{\bibfnamefont{M.}~\bibnamefont{Mai}},
  \bibinfo{author}{\bibfnamefont{M.}~\bibnamefont{Petschlies}},
  \bibinfo{author}{\bibfnamefont{F.}~\bibnamefont{Pittler}},
  \bibinfo{author}{\bibfnamefont{M.}~\bibnamefont{Ueding}},
  \bibinfo{author}{\bibfnamefont{C.}~\bibnamefont{Urbach}}, \bibnamefont{and}
  \bibinfo{author}{\bibfnamefont{M.}~\bibnamefont{Werner}}
  (\bibinfo{collaboration}{ETM}) (\bibinfo{year}{2020}), \eprint{2006.13805}.

\bibitem[{\citenamefont{Aoyama et~al.}(2020)}]{Aoyama:2020ynm}
\bibinfo{author}{\bibfnamefont{T.}~\bibnamefont{Aoyama}} \bibnamefont{et~al.},
  \bibinfo{journal}{Phys. Rept.} \textbf{\bibinfo{volume}{887}},
  \bibinfo{pages}{1} (\bibinfo{year}{2020}), \eprint{2006.04822}.

\bibitem[{\citenamefont{Colangelo et~al.}(2019)\citenamefont{Colangelo,
  Hoferichter, and Stoffer}}]{Colangelo:2018mtw}
\bibinfo{author}{\bibfnamefont{G.}~\bibnamefont{Colangelo}},
  \bibinfo{author}{\bibfnamefont{M.}~\bibnamefont{Hoferichter}},
  \bibnamefont{and} \bibinfo{author}{\bibfnamefont{P.}~\bibnamefont{Stoffer}},
  \bibinfo{journal}{JHEP} \textbf{\bibinfo{volume}{02}}, \bibinfo{pages}{006}
  (\bibinfo{year}{2019}), \eprint{1810.00007}.

\bibitem[{\citenamefont{Ananthanarayan
  et~al.}(2018)\citenamefont{Ananthanarayan, Caprini, and
  Das}}]{Ananthanarayan:2018nyx}
\bibinfo{author}{\bibfnamefont{B.}~\bibnamefont{Ananthanarayan}},
  \bibinfo{author}{\bibfnamefont{I.}~\bibnamefont{Caprini}}, \bibnamefont{and}
  \bibinfo{author}{\bibfnamefont{D.}~\bibnamefont{Das}},
  \bibinfo{journal}{Phys. Rev. D} \textbf{\bibinfo{volume}{98}},
  \bibinfo{pages}{114015} (\bibinfo{year}{2018}), \eprint{1810.09265}.

\bibitem[{\citenamefont{Davier et~al.}(2020)\citenamefont{Davier, Hoecker,
  Malaescu, and Zhang}}]{Davier:2019can}
\bibinfo{author}{\bibfnamefont{M.}~\bibnamefont{Davier}},
  \bibinfo{author}{\bibfnamefont{A.}~\bibnamefont{Hoecker}},
  \bibinfo{author}{\bibfnamefont{B.}~\bibnamefont{Malaescu}}, \bibnamefont{and}
  \bibinfo{author}{\bibfnamefont{Z.}~\bibnamefont{Zhang}},
  \bibinfo{journal}{Eur. Phys. J.} \textbf{\bibinfo{volume}{C80}},
  \bibinfo{pages}{241} (\bibinfo{year}{2020}), \bibinfo{note}{[Erratum: Eur.
  Phys. J. {\bf C80}, 410 (2020)]}, \eprint{1908.00921}.

\bibitem[{\citenamefont{Keshavarzi et~al.}(2020)\citenamefont{Keshavarzi,
  Nomura, and Teubner}}]{keshavarzi:2019abf}
\bibinfo{author}{\bibfnamefont{A.}~\bibnamefont{Keshavarzi}},
  \bibinfo{author}{\bibfnamefont{D.}~\bibnamefont{Nomura}}, \bibnamefont{and}
  \bibinfo{author}{\bibfnamefont{T.}~\bibnamefont{Teubner}},
  \bibinfo{journal}{Phys. Rev.} \textbf{\bibinfo{volume}{D101}},
  \bibinfo{pages}{014029} (\bibinfo{year}{2020}), \eprint{1911.00367}.

\bibitem[{\citenamefont{Colangelo et~al.}(2020)\citenamefont{Colangelo,
  Hoferichter, and Stoffer}}]{Colangelo:2020lcg}
\bibinfo{author}{\bibfnamefont{G.}~\bibnamefont{Colangelo}},
  \bibinfo{author}{\bibfnamefont{M.}~\bibnamefont{Hoferichter}},
  \bibnamefont{and} \bibinfo{author}{\bibfnamefont{P.}~\bibnamefont{Stoffer}}
  (\bibinfo{year}{2020}), \eprint{2010.07943}.

\bibitem[{\citenamefont{Belushkin et~al.}(2007)\citenamefont{Belushkin, Hammer,
  and Mei{\ss}ner}}]{Belushkin:2006qa}
\bibinfo{author}{\bibfnamefont{M.~A.} \bibnamefont{Belushkin}},
  \bibinfo{author}{\bibfnamefont{H.-W.} \bibnamefont{Hammer}},
  \bibnamefont{and} \bibinfo{author}{\bibfnamefont{U.-G.}
  \bibnamefont{Mei{\ss}ner}}, \bibinfo{journal}{Phys. Rev. C}
  \textbf{\bibinfo{volume}{75}}, \bibinfo{pages}{035202}
  (\bibinfo{year}{2007}), \eprint{hep-ph/0608337}.

\bibitem[{\citenamefont{Lorenz et~al.}(2015)\citenamefont{Lorenz, Mei{\ss}ner,
  Hammer, and Dong}}]{Lorenz:2014yda}
\bibinfo{author}{\bibfnamefont{I.~T.} \bibnamefont{Lorenz}},
  \bibinfo{author}{\bibfnamefont{U.-G.} \bibnamefont{Mei{\ss}ner}},
  \bibinfo{author}{\bibfnamefont{H.-W.} \bibnamefont{Hammer}},
  \bibnamefont{and} \bibinfo{author}{\bibfnamefont{Y.~B.} \bibnamefont{Dong}},
  \bibinfo{journal}{Phys. Rev. D} \textbf{\bibinfo{volume}{91}},
  \bibinfo{pages}{014023} (\bibinfo{year}{2015}), \eprint{1411.1704}.

\bibitem[{\citenamefont{Hoferichter et~al.}(2016)\citenamefont{Hoferichter,
  Kubis, Ruiz~de Elvira, Hammer, and Mei{\ss}ner}}]{Hoferichter:2016duk}
\bibinfo{author}{\bibfnamefont{M.}~\bibnamefont{Hoferichter}},
  \bibinfo{author}{\bibfnamefont{B.}~\bibnamefont{Kubis}},
  \bibinfo{author}{\bibfnamefont{J.}~\bibnamefont{Ruiz~de Elvira}},
  \bibinfo{author}{\bibfnamefont{H.-W.} \bibnamefont{Hammer}},
  \bibnamefont{and} \bibinfo{author}{\bibfnamefont{U.-G.}
  \bibnamefont{Mei{\ss}ner}}, \bibinfo{journal}{Eur. Phys. J. A}
  \textbf{\bibinfo{volume}{52}}, \bibinfo{pages}{331} (\bibinfo{year}{2016}),
  \eprint{1609.06722}.

\bibitem[{\citenamefont{Hoferichter et~al.}(2019)\citenamefont{Hoferichter,
  Kubis, Ruiz~de Elvira, and Stoffer}}]{Hoferichter:2018zwu}
\bibinfo{author}{\bibfnamefont{M.}~\bibnamefont{Hoferichter}},
  \bibinfo{author}{\bibfnamefont{B.}~\bibnamefont{Kubis}},
  \bibinfo{author}{\bibfnamefont{J.}~\bibnamefont{Ruiz~de Elvira}},
  \bibnamefont{and} \bibinfo{author}{\bibfnamefont{P.}~\bibnamefont{Stoffer}},
  \bibinfo{journal}{Phys. Rev. Lett.} \textbf{\bibinfo{volume}{122}},
  \bibinfo{pages}{122001} (\bibinfo{year}{2019}), \bibinfo{note}{[Erratum:
  Phys. Rev. Lett. \textbf{124}, 199901 (2020)]}, \eprint{1811.11181}.

\bibitem[{\citenamefont{Hoferichter et~al.}(2012)\citenamefont{Hoferichter,
  Kubis, and Sakkas}}]{Hoferichter:2012pm}
\bibinfo{author}{\bibfnamefont{M.}~\bibnamefont{Hoferichter}},
  \bibinfo{author}{\bibfnamefont{B.}~\bibnamefont{Kubis}}, \bibnamefont{and}
  \bibinfo{author}{\bibfnamefont{D.}~\bibnamefont{Sakkas}},
  \bibinfo{journal}{Phys. Rev. D} \textbf{\bibinfo{volume}{86}},
  \bibinfo{pages}{116009} (\bibinfo{year}{2012}), \eprint{1210.6793}.

\bibitem[{\citenamefont{Hoferichter et~al.}(2017)\citenamefont{Hoferichter,
  Kubis, and Zanke}}]{Hoferichter:2017ftn}
\bibinfo{author}{\bibfnamefont{M.}~\bibnamefont{Hoferichter}},
  \bibinfo{author}{\bibfnamefont{B.}~\bibnamefont{Kubis}}, \bibnamefont{and}
  \bibinfo{author}{\bibfnamefont{M.}~\bibnamefont{Zanke}},
  \bibinfo{journal}{Phys. Rev. D} \textbf{\bibinfo{volume}{96}},
  \bibinfo{pages}{114016} (\bibinfo{year}{2017}), \eprint{1710.00824}.

\bibitem[{\citenamefont{Brice\~no et~al.}(2015)\citenamefont{Brice\~no, Dudek,
  Edwards, Shultz, Thomas, and Wilson}}]{Briceno:2015dca}
\bibinfo{author}{\bibfnamefont{R.~A.} \bibnamefont{Brice\~no}},
  \bibinfo{author}{\bibfnamefont{J.~J.} \bibnamefont{Dudek}},
  \bibinfo{author}{\bibfnamefont{R.~G.} \bibnamefont{Edwards}},
  \bibinfo{author}{\bibfnamefont{C.~J.} \bibnamefont{Shultz}},
  \bibinfo{author}{\bibfnamefont{C.~E.} \bibnamefont{Thomas}},
  \bibnamefont{and} \bibinfo{author}{\bibfnamefont{D.~J.}
  \bibnamefont{Wilson}}, \bibinfo{journal}{Phys. Rev. Lett.}
  \textbf{\bibinfo{volume}{115}}, \bibinfo{pages}{242001}
  (\bibinfo{year}{2015}), \eprint{1507.06622}.

\bibitem[{\citenamefont{Brice\~no et~al.}(2016)\citenamefont{Brice\~no, Dudek,
  Edwards, Shultz, Thomas, and Wilson}}]{Briceno:2016kkp}
\bibinfo{author}{\bibfnamefont{R.~A.} \bibnamefont{Brice\~no}},
  \bibinfo{author}{\bibfnamefont{J.~J.} \bibnamefont{Dudek}},
  \bibinfo{author}{\bibfnamefont{R.~G.} \bibnamefont{Edwards}},
  \bibinfo{author}{\bibfnamefont{C.~J.} \bibnamefont{Shultz}},
  \bibinfo{author}{\bibfnamefont{C.~E.} \bibnamefont{Thomas}},
  \bibnamefont{and} \bibinfo{author}{\bibfnamefont{D.~J.}
  \bibnamefont{Wilson}}, \bibinfo{journal}{Phys. Rev. D}
  \textbf{\bibinfo{volume}{93}}, \bibinfo{pages}{114508}
  (\bibinfo{year}{2016}), \eprint{1604.03530}.

\bibitem[{\citenamefont{Alexandrou et~al.}(2018)\citenamefont{Alexandrou,
  Leskovec, Meinel, Negele, Paul, Petschlies, Pochinsky, Rendon, and
  Syritsyn}}]{Alexandrou:2018jbt}
\bibinfo{author}{\bibfnamefont{C.}~\bibnamefont{Alexandrou}},
  \bibinfo{author}{\bibfnamefont{L.}~\bibnamefont{Leskovec}},
  \bibinfo{author}{\bibfnamefont{S.}~\bibnamefont{Meinel}},
  \bibinfo{author}{\bibfnamefont{J.}~\bibnamefont{Negele}},
  \bibinfo{author}{\bibfnamefont{S.}~\bibnamefont{Paul}},
  \bibinfo{author}{\bibfnamefont{M.}~\bibnamefont{Petschlies}},
  \bibinfo{author}{\bibfnamefont{A.}~\bibnamefont{Pochinsky}},
  \bibinfo{author}{\bibfnamefont{G.}~\bibnamefont{Rendon}}, \bibnamefont{and}
  \bibinfo{author}{\bibfnamefont{S.}~\bibnamefont{Syritsyn}},
  \bibinfo{journal}{Phys. Rev. D} \textbf{\bibinfo{volume}{98}},
  \bibinfo{pages}{074502} (\bibinfo{year}{2018}), \eprint{1807.08357}.

\bibitem[{\citenamefont{Niehus et~al.}(2019)\citenamefont{Niehus, Hoferichter,
  and Kubis}}]{Niehus:2019nkl}
\bibinfo{author}{\bibfnamefont{M.}~\bibnamefont{Niehus}},
  \bibinfo{author}{\bibfnamefont{M.}~\bibnamefont{Hoferichter}},
  \bibnamefont{and} \bibinfo{author}{\bibfnamefont{B.}~\bibnamefont{Kubis}},
  \bibinfo{journal}{PoS} \textbf{\bibinfo{volume}{CD2018}},
  \bibinfo{pages}{076} (\bibinfo{year}{2019}), \eprint{1902.10150}.

\bibitem[{\citenamefont{Dax et~al.}(2018)\citenamefont{Dax, Isken, and
  Kubis}}]{Dax:2018rvs}
\bibinfo{author}{\bibfnamefont{M.}~\bibnamefont{Dax}},
  \bibinfo{author}{\bibfnamefont{T.}~\bibnamefont{Isken}}, \bibnamefont{and}
  \bibinfo{author}{\bibfnamefont{B.}~\bibnamefont{Kubis}},
  \bibinfo{journal}{Eur. Phys. J. C} \textbf{\bibinfo{volume}{78}},
  \bibinfo{pages}{859} (\bibinfo{year}{2018}), \eprint{1808.08957}.

\bibitem[{\citenamefont{Truong}(1988)}]{Truong:1988zp}
\bibinfo{author}{\bibfnamefont{T.~N.} \bibnamefont{Truong}},
  \bibinfo{journal}{Phys. Rev. Lett.} \textbf{\bibinfo{volume}{61}},
  \bibinfo{pages}{2526} (\bibinfo{year}{1988}).

\bibitem[{\citenamefont{Dobado et~al.}(1990)\citenamefont{Dobado, Herrero, and
  Truong}}]{Dobado:1989qm}
\bibinfo{author}{\bibfnamefont{A.}~\bibnamefont{Dobado}},
  \bibinfo{author}{\bibfnamefont{M.~J.} \bibnamefont{Herrero}},
  \bibnamefont{and} \bibinfo{author}{\bibfnamefont{T.~N.}
  \bibnamefont{Truong}}, \bibinfo{journal}{Phys. Lett. B}
  \textbf{\bibinfo{volume}{235}}, \bibinfo{pages}{134} (\bibinfo{year}{1990}).

\bibitem[{\citenamefont{Truong}(1991)}]{Truong1991}
\bibinfo{author}{\bibfnamefont{T.~N.} \bibnamefont{Truong}},
  \bibinfo{journal}{Phys. Rev. Lett.} \textbf{\bibinfo{volume}{67}},
  \bibinfo{pages}{2260} (\bibinfo{year}{1991}).

\bibitem[{\citenamefont{Dobado and Pel{\'{a}}ez}(1993)}]{Dobado1993}
\bibinfo{author}{\bibfnamefont{A.}~\bibnamefont{Dobado}} \bibnamefont{and}
  \bibinfo{author}{\bibfnamefont{J.~R.} \bibnamefont{Pel{\'{a}}ez}},
  \bibinfo{journal}{Phys. Rev. D} \textbf{\bibinfo{volume}{47}},
  \bibinfo{pages}{4883} (\bibinfo{year}{1993}), \eprint{hep-ph/9301276}.

\bibitem[{\citenamefont{Dobado and Pel{\'{a}}ez}(1997)}]{Dobado1997}
\bibinfo{author}{\bibfnamefont{A.}~\bibnamefont{Dobado}} \bibnamefont{and}
  \bibinfo{author}{\bibfnamefont{J.~R.} \bibnamefont{Pel{\'{a}}ez}},
  \bibinfo{journal}{Phys. Rev. D} \textbf{\bibinfo{volume}{56}},
  \bibinfo{pages}{3057} (\bibinfo{year}{1997}), \eprint{hep-ph/9604416}.

\bibitem[{\citenamefont{Guerrero and Oller}(1999)}]{Guerrero:1998ei}
\bibinfo{author}{\bibfnamefont{F.}~\bibnamefont{Guerrero}} \bibnamefont{and}
  \bibinfo{author}{\bibfnamefont{J.~A.} \bibnamefont{Oller}},
  \bibinfo{journal}{Nucl. Phys. B} \textbf{\bibinfo{volume}{537}},
  \bibinfo{pages}{459} (\bibinfo{year}{1999}), \bibinfo{note}{[Erratum: Nucl.
  Phys. B {\bf 602}, 641 (2001)]}, \eprint{hep-ph/9805334}.

\bibitem[{\citenamefont{G{\'o}mez~Nicola and
  Pel{\'a}ez}(2002)}]{GomezNicola:2001as}
\bibinfo{author}{\bibfnamefont{A.}~\bibnamefont{G{\'o}mez~Nicola}}
  \bibnamefont{and} \bibinfo{author}{\bibfnamefont{J.~R.}
  \bibnamefont{Pel{\'a}ez}}, \bibinfo{journal}{Phys. Rev. D}
  \textbf{\bibinfo{volume}{65}}, \bibinfo{pages}{054009}
  (\bibinfo{year}{2002}), \eprint{hep-ph/0109056}.

\bibitem[{\citenamefont{Nieves et~al.}(2002)\citenamefont{Nieves,
  Pav{\'o}n~Valderrama, and Ruiz~Arriola}}]{Nieves:2001de}
\bibinfo{author}{\bibfnamefont{J.}~\bibnamefont{Nieves}},
  \bibinfo{author}{\bibfnamefont{M.}~\bibnamefont{Pav{\'o}n~Valderrama}},
  \bibnamefont{and}
  \bibinfo{author}{\bibfnamefont{E.}~\bibnamefont{Ruiz~Arriola}},
  \bibinfo{journal}{Phys. Rev. D} \textbf{\bibinfo{volume}{65}},
  \bibinfo{pages}{036002} (\bibinfo{year}{2002}), \eprint{hep-ph/0109077}.

\bibitem[{\citenamefont{Dobado and Pel{\'a}ez}(2002)}]{Dobado:2001rv}
\bibinfo{author}{\bibfnamefont{A.}~\bibnamefont{Dobado}} \bibnamefont{and}
  \bibinfo{author}{\bibfnamefont{J.~R.} \bibnamefont{Pel{\'a}ez}},
  \bibinfo{journal}{Phys. Rev. D} \textbf{\bibinfo{volume}{65}},
  \bibinfo{pages}{077502} (\bibinfo{year}{2002}), \eprint{hep-ph/0111140}.

\bibitem[{\citenamefont{G{\'o}mez~Nicola
  et~al.}(2008)\citenamefont{G{\'o}mez~Nicola, Pel{\'a}ez, and
  R{\'i}os}}]{GomezNicola:2007qj}
\bibinfo{author}{\bibfnamefont{A.}~\bibnamefont{G{\'o}mez~Nicola}},
  \bibinfo{author}{\bibfnamefont{J.~R.} \bibnamefont{Pel{\'a}ez}},
  \bibnamefont{and} \bibinfo{author}{\bibfnamefont{G.}~\bibnamefont{R{\'i}os}},
  \bibinfo{journal}{Phys. Rev. D} \textbf{\bibinfo{volume}{77}},
  \bibinfo{pages}{056006} (\bibinfo{year}{2008}), \eprint{0712.2763}.

\bibitem[{\citenamefont{Bijnens et~al.}(1996)\citenamefont{Bijnens, Colangelo,
  Ecker, Gasser, and Sainio}}]{Bijnens1996}
\bibinfo{author}{\bibfnamefont{J.}~\bibnamefont{Bijnens}},
  \bibinfo{author}{\bibfnamefont{G.}~\bibnamefont{Colangelo}},
  \bibinfo{author}{\bibfnamefont{G.}~\bibnamefont{Ecker}},
  \bibinfo{author}{\bibfnamefont{J.}~\bibnamefont{Gasser}}, \bibnamefont{and}
  \bibinfo{author}{\bibfnamefont{M.~E.} \bibnamefont{Sainio}},
  \bibinfo{journal}{Phys. Lett. B} \textbf{\bibinfo{volume}{374}},
  \bibinfo{pages}{210} (\bibinfo{year}{1996}), \eprint{hep-ph/9511397}.

\bibitem[{\citenamefont{Hanhart et~al.}(2008)\citenamefont{Hanhart, Pel\'aez,
  and R\'ios}}]{Hanhart:2008mx}
\bibinfo{author}{\bibfnamefont{C.}~\bibnamefont{Hanhart}},
  \bibinfo{author}{\bibfnamefont{J.~R.} \bibnamefont{Pel\'aez}},
  \bibnamefont{and} \bibinfo{author}{\bibfnamefont{G.}~\bibnamefont{R\'ios}},
  \bibinfo{journal}{Phys. Rev. Lett.} \textbf{\bibinfo{volume}{100}},
  \bibinfo{pages}{152001} (\bibinfo{year}{2008}), \eprint{0801.2871}.

\bibitem[{\citenamefont{Nebreda and Pel\'aez}(2010)}]{Nebreda:2010wv}
\bibinfo{author}{\bibfnamefont{J.}~\bibnamefont{Nebreda}} \bibnamefont{and}
  \bibinfo{author}{\bibfnamefont{J.~R.} \bibnamefont{Pel\'aez}},
  \bibinfo{journal}{Phys. Rev.} \textbf{\bibinfo{volume}{D81}},
  \bibinfo{pages}{054035} (\bibinfo{year}{2010}), \eprint{1001.5237}.

\bibitem[{\citenamefont{Pel\'aez and R\'ios}(2010)}]{Pelaez:2010fj}
\bibinfo{author}{\bibfnamefont{J.~R.} \bibnamefont{Pel\'aez}} \bibnamefont{and}
  \bibinfo{author}{\bibfnamefont{G.}~\bibnamefont{R\'ios}},
  \bibinfo{journal}{Phys. Rev. D} \textbf{\bibinfo{volume}{82}},
  \bibinfo{pages}{114002} (\bibinfo{year}{2010}), \eprint{1010.6008}.

\bibitem[{\citenamefont{Nebreda et~al.}(2011)\citenamefont{Nebreda, Pel\'aez,
  and R\'ios}}]{Nebreda:2011di}
\bibinfo{author}{\bibfnamefont{J.}~\bibnamefont{Nebreda}},
  \bibinfo{author}{\bibfnamefont{J.~R.} \bibnamefont{Pel\'aez}},
  \bibnamefont{and} \bibinfo{author}{\bibfnamefont{G.}~\bibnamefont{R\'ios}},
  \bibinfo{journal}{Phys. Rev. D} \textbf{\bibinfo{volume}{83}},
  \bibinfo{pages}{094011} (\bibinfo{year}{2011}), \eprint{1101.2171}.

\bibitem[{\citenamefont{Bolton et~al.}(2016)\citenamefont{Bolton, Brice\~no,
  and Wilson}}]{Bolton:2015psa}
\bibinfo{author}{\bibfnamefont{D.~R.} \bibnamefont{Bolton}},
  \bibinfo{author}{\bibfnamefont{R.~A.} \bibnamefont{Brice\~no}},
  \bibnamefont{and} \bibinfo{author}{\bibfnamefont{D.~J.}
  \bibnamefont{Wilson}}, \bibinfo{journal}{Phys. Lett. B}
  \textbf{\bibinfo{volume}{757}}, \bibinfo{pages}{50} (\bibinfo{year}{2016}),
  \eprint{1507.07928}.

\bibitem[{\citenamefont{Hu et~al.}(2016)\citenamefont{Hu, Molina, D{\"o}ring,
  and Alexandru}}]{Hu:2016shf}
\bibinfo{author}{\bibfnamefont{B.}~\bibnamefont{Hu}},
  \bibinfo{author}{\bibfnamefont{R.}~\bibnamefont{Molina}},
  \bibinfo{author}{\bibfnamefont{M.}~\bibnamefont{D{\"o}ring}},
  \bibnamefont{and}
  \bibinfo{author}{\bibfnamefont{A.}~\bibnamefont{Alexandru}},
  \bibinfo{journal}{Phys. Rev. Lett.} \textbf{\bibinfo{volume}{117}},
  \bibinfo{pages}{122001} (\bibinfo{year}{2016}), \eprint{1605.04823}.

\bibitem[{\citenamefont{D{\"o}ring et~al.}(2018)\citenamefont{D{\"o}ring, Hu,
  and Mai}}]{Doring:2016bdr}
\bibinfo{author}{\bibfnamefont{M.}~\bibnamefont{D{\"o}ring}},
  \bibinfo{author}{\bibfnamefont{B.}~\bibnamefont{Hu}}, \bibnamefont{and}
  \bibinfo{author}{\bibfnamefont{M.}~\bibnamefont{Mai}},
  \bibinfo{journal}{Phys. Lett. B} \textbf{\bibinfo{volume}{782}},
  \bibinfo{pages}{785} (\bibinfo{year}{2018}), \eprint{1610.10070}.

\bibitem[{\citenamefont{Hu et~al.}(2017)\citenamefont{Hu, Molina, D{\"o}ring,
  Mai, and Alexandru}}]{Hu:2017wli}
\bibinfo{author}{\bibfnamefont{B.}~\bibnamefont{Hu}},
  \bibinfo{author}{\bibfnamefont{R.}~\bibnamefont{Molina}},
  \bibinfo{author}{\bibfnamefont{M.}~\bibnamefont{D{\"o}ring}},
  \bibinfo{author}{\bibfnamefont{M.}~\bibnamefont{Mai}}, \bibnamefont{and}
  \bibinfo{author}{\bibfnamefont{A.}~\bibnamefont{Alexandru}},
  \bibinfo{journal}{Phys. Rev. D} \textbf{\bibinfo{volume}{96}},
  \bibinfo{pages}{034520} (\bibinfo{year}{2017}), \eprint{1704.06248}.

\bibitem[{\citenamefont{Molina and Ruiz~de Elvira}(2020)}]{Molina:2020qpw}
\bibinfo{author}{\bibfnamefont{R.}~\bibnamefont{Molina}} \bibnamefont{and}
  \bibinfo{author}{\bibfnamefont{J.}~\bibnamefont{Ruiz~de Elvira}},
  \bibinfo{journal}{JHEP} \textbf{\bibinfo{volume}{11}}, \bibinfo{pages}{017}
  (\bibinfo{year}{2020}), \eprint{2005.13584}.

\bibitem[{Sup()}]{Supp}
\bibinfo{note}{The ChPT expressions are also provided in the form of a
  \texttt{Mathematica} notebook.}

\bibitem[{\citenamefont{Virtanen et~al.}(2020)\citenamefont{Virtanen, Gommers,
  Oliphant et~al.}}]{2020SciPy-NMeth}
\bibinfo{author}{\bibfnamefont{P.}~\bibnamefont{Virtanen}},
  \bibinfo{author}{\bibfnamefont{R.}~\bibnamefont{Gommers}},
  \bibinfo{author}{\bibfnamefont{T.~E.} \bibnamefont{Oliphant}},
  \bibnamefont{et~al.}, \bibinfo{journal}{Nature Methods}
  \textbf{\bibinfo{volume}{17}}, \bibinfo{pages}{261} (\bibinfo{year}{2020}).

\bibitem[{\citenamefont{Edwards et~al.}(2011)\citenamefont{Edwards, Dudek,
  Richards, and Wallace}}]{Edwards:2011jj}
\bibinfo{author}{\bibfnamefont{R.~G.} \bibnamefont{Edwards}},
  \bibinfo{author}{\bibfnamefont{J.~J.} \bibnamefont{Dudek}},
  \bibinfo{author}{\bibfnamefont{D.~G.} \bibnamefont{Richards}},
  \bibnamefont{and} \bibinfo{author}{\bibfnamefont{S.~J.}
  \bibnamefont{Wallace}}, \bibinfo{journal}{Phys. Rev. D}
  \textbf{\bibinfo{volume}{84}}, \bibinfo{pages}{074508}
  (\bibinfo{year}{2011}), \eprint{1104.5152}.

\bibitem[{\citenamefont{Bijnens et~al.}(2000)\citenamefont{Bijnens, Colangelo,
  and Ecker}}]{Bijnens:1999hw}
\bibinfo{author}{\bibfnamefont{J.}~\bibnamefont{Bijnens}},
  \bibinfo{author}{\bibfnamefont{G.}~\bibnamefont{Colangelo}},
  \bibnamefont{and} \bibinfo{author}{\bibfnamefont{G.}~\bibnamefont{Ecker}},
  \bibinfo{journal}{Annals Phys.} \textbf{\bibinfo{volume}{280}},
  \bibinfo{pages}{100} (\bibinfo{year}{2000}), \eprint{hep-ph/9907333}.

\bibitem[{\citenamefont{Weinberg}(1966)}]{Weinberg:1966kf}
\bibinfo{author}{\bibfnamefont{S.}~\bibnamefont{Weinberg}},
  \bibinfo{journal}{Phys. Rev. Lett.} \textbf{\bibinfo{volume}{17}},
  \bibinfo{pages}{616} (\bibinfo{year}{1966}).

\bibitem[{\citenamefont{L{\"u}scher}(1991)}]{Luscher:1990ux}
\bibinfo{author}{\bibfnamefont{M.}~\bibnamefont{L{\"u}scher}},
  \bibinfo{journal}{Nucl. Phys.} \textbf{\bibinfo{volume}{B354}},
  \bibinfo{pages}{531} (\bibinfo{year}{1991}).

\bibitem[{\citenamefont{Bruno et~al.}(2017)\citenamefont{Bruno, Korzec, and
  Schaefer}}]{Bruno:2016plf}
\bibinfo{author}{\bibfnamefont{M.}~\bibnamefont{Bruno}},
  \bibinfo{author}{\bibfnamefont{T.}~\bibnamefont{Korzec}}, \bibnamefont{and}
  \bibinfo{author}{\bibfnamefont{S.}~\bibnamefont{Schaefer}},
  \bibinfo{journal}{Phys. Rev. D} \textbf{\bibinfo{volume}{95}},
  \bibinfo{pages}{074504} (\bibinfo{year}{2017}), \eprint{1608.08900}.

\bibitem[{\citenamefont{Dudek et~al.}(2013)\citenamefont{Dudek, Edwards, and
  Thomas}}]{Dudek2013}
\bibinfo{author}{\bibfnamefont{J.~J.} \bibnamefont{Dudek}},
  \bibinfo{author}{\bibfnamefont{R.~G.} \bibnamefont{Edwards}},
  \bibnamefont{and} \bibinfo{author}{\bibfnamefont{C.~E.} \bibnamefont{Thomas}}
  (\bibinfo{collaboration}{Hadron Spectrum}), \bibinfo{journal}{Phys. Rev. D}
  \textbf{\bibinfo{volume}{87}}, \bibinfo{pages}{034505}
  (\bibinfo{year}{2013}), \bibinfo{note}{[Erratum: Phys. Rev. D {\bf 90},
  099902 (2014)]}, \eprint{1212.0830}.

\bibitem[{\citenamefont{Gasser et~al.}(2007)\citenamefont{Gasser, Haefeli,
  Ivanov, and Schmid}}]{Gasser:2007sg}
\bibinfo{author}{\bibfnamefont{J.}~\bibnamefont{Gasser}},
  \bibinfo{author}{\bibfnamefont{C.}~\bibnamefont{Haefeli}},
  \bibinfo{author}{\bibfnamefont{M.~A.} \bibnamefont{Ivanov}},
  \bibnamefont{and} \bibinfo{author}{\bibfnamefont{M.}~\bibnamefont{Schmid}},
  \bibinfo{journal}{Phys. Lett. B} \textbf{\bibinfo{volume}{652}},
  \bibinfo{pages}{21} (\bibinfo{year}{2007}), \eprint{0706.0955}.

\bibitem[{\citenamefont{Gasser et~al.}(2009)\citenamefont{Gasser, Haefeli,
  Ivanov, and Schmid}}]{Gasser:2009hr}
\bibinfo{author}{\bibfnamefont{J.}~\bibnamefont{Gasser}},
  \bibinfo{author}{\bibfnamefont{C.}~\bibnamefont{Haefeli}},
  \bibinfo{author}{\bibfnamefont{M.~A.} \bibnamefont{Ivanov}},
  \bibnamefont{and} \bibinfo{author}{\bibfnamefont{M.}~\bibnamefont{Schmid}},
  \bibinfo{journal}{Phys. Lett. B} \textbf{\bibinfo{volume}{675}},
  \bibinfo{pages}{49} (\bibinfo{year}{2009}), \eprint{0903.0801}.

\bibitem[{\citenamefont{Storn and Price}(1997)}]{Storn1997}
\bibinfo{author}{\bibfnamefont{R.}~\bibnamefont{Storn}} \bibnamefont{and}
  \bibinfo{author}{\bibfnamefont{K.}~\bibnamefont{Price}},
  \bibinfo{journal}{Journal of Global Optimization}
  \textbf{\bibinfo{volume}{11}}, \bibinfo{pages}{341} (\bibinfo{year}{1997}).

\bibitem[{\citenamefont{Press et~al.}(2007)\citenamefont{Press, Teukolsky,
  Vetterling, and Flannery}}]{Press:2007num}
\bibinfo{author}{\bibfnamefont{W.~H.} \bibnamefont{Press}},
  \bibinfo{author}{\bibfnamefont{S.~A.} \bibnamefont{Teukolsky}},
  \bibinfo{author}{\bibfnamefont{W.~T.} \bibnamefont{Vetterling}},
  \bibnamefont{and} \bibinfo{author}{\bibfnamefont{B.~P.}
  \bibnamefont{Flannery}}, \emph{\bibinfo{title}{Numerical Recipes: The Art of
  Scientific Computing}} (\bibinfo{publisher}{Cambridge University Press},
  \bibinfo{address}{USA}, \bibinfo{year}{2007}), \bibinfo{edition}{3rd} ed.,
  ISBN \bibinfo{isbn}{9780521880688}.

\bibitem[{\citenamefont{Epelbaum et~al.}(2015)\citenamefont{Epelbaum, Krebs,
  and Mei{\ss}ner}}]{Epelbaum:2014efa}
\bibinfo{author}{\bibfnamefont{E.}~\bibnamefont{Epelbaum}},
  \bibinfo{author}{\bibfnamefont{H.}~\bibnamefont{Krebs}}, \bibnamefont{and}
  \bibinfo{author}{\bibfnamefont{U.-G.} \bibnamefont{Mei{\ss}ner}},
  \bibinfo{journal}{Eur. Phys. J. A} \textbf{\bibinfo{volume}{51}},
  \bibinfo{pages}{53} (\bibinfo{year}{2015}), \eprint{1412.0142}.

\bibitem[{\citenamefont{Bijnens and Ecker}(2014)}]{Bijnens:2014lea}
\bibinfo{author}{\bibfnamefont{J.}~\bibnamefont{Bijnens}} \bibnamefont{and}
  \bibinfo{author}{\bibfnamefont{G.}~\bibnamefont{Ecker}},
  \bibinfo{journal}{Ann. Rev. Nucl. Part. Sci.} \textbf{\bibinfo{volume}{64}},
  \bibinfo{pages}{149} (\bibinfo{year}{2014}), \eprint{1405.6488}.

\bibitem[{\citenamefont{Aoki et~al.}(2020)}]{Aoki:2019cca}
\bibinfo{author}{\bibfnamefont{S.}~\bibnamefont{Aoki}} \bibnamefont{et~al.}
  (\bibinfo{collaboration}{Flavour Lattice Averaging Group}),
  \bibinfo{journal}{Eur. Phys. J. C} \textbf{\bibinfo{volume}{80}},
  \bibinfo{pages}{113} (\bibinfo{year}{2020}), \eprint{1902.08191}.

\bibitem[{\citenamefont{Bazavov et~al.}(2010)}]{Bazavov:2010hj}
\bibinfo{author}{\bibfnamefont{A.}~\bibnamefont{Bazavov}} \bibnamefont{et~al.}
  (\bibinfo{collaboration}{MILC}), \bibinfo{journal}{PoS}
  \textbf{\bibinfo{volume}{LATTICE2010}}, \bibinfo{pages}{074}
  (\bibinfo{year}{2010}), \eprint{1012.0868}.

\bibitem[{\citenamefont{Beane et~al.}(2012)\citenamefont{Beane, Detmold,
  Junnarkar, Luu, Orginos, Parre{\~{n}}o, Savage, Torok, and
  Walker-Loud}}]{Beane:2011zm}
\bibinfo{author}{\bibfnamefont{S.~R.} \bibnamefont{Beane}},
  \bibinfo{author}{\bibfnamefont{W.}~\bibnamefont{Detmold}},
  \bibinfo{author}{\bibfnamefont{P.~M.} \bibnamefont{Junnarkar}},
  \bibinfo{author}{\bibfnamefont{T.~C.} \bibnamefont{Luu}},
  \bibinfo{author}{\bibfnamefont{K.}~\bibnamefont{Orginos}},
  \bibinfo{author}{\bibfnamefont{A.}~\bibnamefont{Parre{\~{n}}o}},
  \bibinfo{author}{\bibfnamefont{M.~J.} \bibnamefont{Savage}},
  \bibinfo{author}{\bibfnamefont{A.}~\bibnamefont{Torok}}, \bibnamefont{and}
  \bibinfo{author}{\bibfnamefont{A.}~\bibnamefont{Walker-Loud}},
  \bibinfo{journal}{Phys. Rev. D} \textbf{\bibinfo{volume}{86}},
  \bibinfo{pages}{094509} (\bibinfo{year}{2012}), \eprint{1108.1380}.

\bibitem[{\citenamefont{Bors\'anyi et~al.}(2013)\citenamefont{Bors\'anyi,
  D{\"u}rr, Fodor, Krieg, Sch{\"a}fer, Scholz, and Szabó}}]{Borsanyi:2012zv}
\bibinfo{author}{\bibfnamefont{S.}~\bibnamefont{Bors\'anyi}},
  \bibinfo{author}{\bibfnamefont{S.}~\bibnamefont{D{\"u}rr}},
  \bibinfo{author}{\bibfnamefont{Z.}~\bibnamefont{Fodor}},
  \bibinfo{author}{\bibfnamefont{S.}~\bibnamefont{Krieg}},
  \bibinfo{author}{\bibfnamefont{A.}~\bibnamefont{Sch{\"a}fer}},
  \bibinfo{author}{\bibfnamefont{E.~E.} \bibnamefont{Scholz}},
  \bibnamefont{and} \bibinfo{author}{\bibfnamefont{K.~K.}
  \bibnamefont{Szabó}}, \bibinfo{journal}{Phys. Rev. D}
  \textbf{\bibinfo{volume}{88}}, \bibinfo{pages}{014513}
  (\bibinfo{year}{2013}), \eprint{1205.0788}.

\bibitem[{\citenamefont{D{\"u}rr et~al.}(2014)}]{Durr:2013goa}
\bibinfo{author}{\bibfnamefont{S.}~\bibnamefont{D{\"u}rr}} \bibnamefont{et~al.}
  (\bibinfo{collaboration}{Budapest-Marseille-Wuppertal}),
  \bibinfo{journal}{Phys. Rev. D} \textbf{\bibinfo{volume}{90}},
  \bibinfo{pages}{114504} (\bibinfo{year}{2014}), \eprint{1310.3626}.

\bibitem[{\citenamefont{Boyle et~al.}(2016)}]{Boyle:2015exm}
\bibinfo{author}{\bibfnamefont{P.~A.} \bibnamefont{Boyle}}
  \bibnamefont{et~al.}, \bibinfo{journal}{Phys. Rev. D}
  \textbf{\bibinfo{volume}{93}}, \bibinfo{pages}{054502}
  (\bibinfo{year}{2016}), \eprint{1511.01950}.

\bibitem[{\citenamefont{Bijnens et~al.}(1997)\citenamefont{Bijnens, Colangelo,
  Ecker, Gasser, and Sainio}}]{Bijnens:1997vq}
\bibinfo{author}{\bibfnamefont{J.}~\bibnamefont{Bijnens}},
  \bibinfo{author}{\bibfnamefont{G.}~\bibnamefont{Colangelo}},
  \bibinfo{author}{\bibfnamefont{G.}~\bibnamefont{Ecker}},
  \bibinfo{author}{\bibfnamefont{J.}~\bibnamefont{Gasser}}, \bibnamefont{and}
  \bibinfo{author}{\bibfnamefont{M.~E.} \bibnamefont{Sainio}},
  \bibinfo{journal}{Nucl. Phys.} \textbf{\bibinfo{volume}{B508}},
  \bibinfo{pages}{263} (\bibinfo{year}{1997}), \bibinfo{note}{[Erratum:
  Nucl.\,Phys.\,{\bf B517},\,639 (1998)]}, \eprint{hep-ph/9707291}.

\bibitem[{\citenamefont{Bijnens et~al.}(1998)\citenamefont{Bijnens, Colangelo,
  and Talavera}}]{Bijnens:1998fm}
\bibinfo{author}{\bibfnamefont{J.}~\bibnamefont{Bijnens}},
  \bibinfo{author}{\bibfnamefont{G.}~\bibnamefont{Colangelo}},
  \bibnamefont{and} \bibinfo{author}{\bibfnamefont{P.}~\bibnamefont{Talavera}},
  \bibinfo{journal}{JHEP} \textbf{\bibinfo{volume}{05}}, \bibinfo{pages}{014}
  (\bibinfo{year}{1998}), \eprint{hep-ph/9805389}.

\bibitem[{\citenamefont{Zyla et~al.}(2020)}]{Zyla:2020zbs}
\bibinfo{author}{\bibfnamefont{P.~A.} \bibnamefont{Zyla}} \bibnamefont{et~al.}
  (\bibinfo{collaboration}{Particle Data Group}), \bibinfo{journal}{PTEP}
  \textbf{\bibinfo{volume}{2020}}, \bibinfo{pages}{083C01}
  (\bibinfo{year}{2020}).

\bibitem[{\citenamefont{Kass and Raftery}(1995)}]{Kass:1995loi}
\bibinfo{author}{\bibfnamefont{R.~E.} \bibnamefont{Kass}} \bibnamefont{and}
  \bibinfo{author}{\bibfnamefont{A.~E.} \bibnamefont{Raftery}},
  \bibinfo{journal}{J. Am. Statist. Assoc.} \textbf{\bibinfo{volume}{90}},
  \bibinfo{pages}{773} (\bibinfo{year}{1995}).

\bibitem[{\citenamefont{Garc\'ia-Mart\'in
  et~al.}(2011)\citenamefont{Garc\'ia-Mart\'in, Kami\'nski, Pel\'aez, and
  Ruiz~de Elvira}}]{GarciaMartin:2011jx}
\bibinfo{author}{\bibfnamefont{R.}~\bibnamefont{Garc\'ia-Mart\'in}},
  \bibinfo{author}{\bibfnamefont{R.}~\bibnamefont{Kami\'nski}},
  \bibinfo{author}{\bibfnamefont{J.~R.} \bibnamefont{Pel\'aez}},
  \bibnamefont{and} \bibinfo{author}{\bibfnamefont{J.}~\bibnamefont{Ruiz~de
  Elvira}}, \bibinfo{journal}{Phys. Rev. Lett.} \textbf{\bibinfo{volume}{107}},
  \bibinfo{pages}{072001} (\bibinfo{year}{2011}), \eprint{1107.1635}.

\bibitem[{\citenamefont{Kawarabayashi and Suzuki}(1966)}]{Kawarabayashi:1966kd}
\bibinfo{author}{\bibfnamefont{K.}~\bibnamefont{Kawarabayashi}}
  \bibnamefont{and} \bibinfo{author}{\bibfnamefont{M.}~\bibnamefont{Suzuki}},
  \bibinfo{journal}{Phys. Rev. Lett.} \textbf{\bibinfo{volume}{16}},
  \bibinfo{pages}{255} (\bibinfo{year}{1966}).

\bibitem[{\citenamefont{Riazuddin and Fayyazuddin}(1966)}]{Riazuddin:1966sw}
\bibinfo{author}{\bibnamefont{Riazuddin}} \bibnamefont{and}
  \bibinfo{author}{\bibnamefont{Fayyazuddin}}, \bibinfo{journal}{Phys. Rev.}
  \textbf{\bibinfo{volume}{147}}, \bibinfo{pages}{1071} (\bibinfo{year}{1966}).

\bibitem[{\citenamefont{Ruiz~de Elvira et~al.}(2017)\citenamefont{Ruiz~de
  Elvira, Mei{\ss}ner, Rusetsky, and Schierholz}}]{RuizdeElvira:2017aet}
\bibinfo{author}{\bibfnamefont{J.}~\bibnamefont{Ruiz~de Elvira}},
  \bibinfo{author}{\bibfnamefont{U.-G.} \bibnamefont{Mei{\ss}ner}},
  \bibinfo{author}{\bibfnamefont{A.}~\bibnamefont{Rusetsky}}, \bibnamefont{and}
  \bibinfo{author}{\bibfnamefont{G.}~\bibnamefont{Schierholz}},
  \bibinfo{journal}{Eur. Phys. J.} \textbf{\bibinfo{volume}{C77}},
  \bibinfo{pages}{659} (\bibinfo{year}{2017}), \eprint{1706.09015}.

\end{thebibliography}

\end{document}